\documentclass[sigconf, nonacm]{acmart}

\usepackage{CJK}
\usepackage[a-1b]{pdfx}

\theoremstyle{definition}
\newtheorem{definition}{Definition}

\usepackage{balance}
\usepackage{multirow}
\usepackage{colortbl}
\usepackage{subfigure}

\usepackage[ruled,linesnumbered]{algorithm2e}

\usepackage{multirow}

\newcommand{\vs}{\textsc{ViSearch}\ }
\newcommand{\visar}{\textsc{Vis-Ar}\ }
\newcommand{\rwf}{\textsc{Rwf}\ }

\newcommand{\ang}[1]{\langle #1 \rangle}
\newcommand{\mydef}{\stackrel{\triangle}{=}}

\newcommand{\tarb}{\mathcal{T}_{\text{\scriptsize ARB}}}
\newcommand{\tvis}{\mathcal{T}_{\text{\scriptsize VIS}}}
\newcommand{\tnotvis}{\mathcal{T}_{\text{\scriptsize NOT-VIS}}}

\newcommand\vldbdoi{XX.XX/XXX.XX}
\newcommand\vldbpages{XXX-XXX}
\newcommand\vldbvolume{14}
\newcommand\vldbissue{1}
\newcommand\vldbyear{2020}
\newcommand\vldbauthors{\authors}
\newcommand\vldbtitle{\shorttitle} 
\newcommand\vldbavailabilityurl{https://github.com/ViSearch/ViSearch}
\newcommand\vldbpagestyle{plain}

\begin{document}

\title{\textsc{ViSearch}: Weak Consistency Measurement for \\ Replicated Data Types}


\author{Lintian Shi,\ Yuqi Zhang,\ $\text{Hengfeng Wei}^{\dagger}$,\ $\text{Yu Huang}^{\dagger}$,\ Xiaoxing Ma}\thanks{$\dagger$ Corresponding authors.}
\affiliation{%
   \department{State Key Laboratory for Novel Software Technology, Nanjing University}
}
\email{lintianshi@smail.nju.edu.cn,\space cs.yqzhang@gmail.com,\space \{hfwei,yuhuang,xxm\}@nju.edu.cn}





\begin{abstract}

 Large-scale replicated data type stores often resort to eventual consistency to guarantee low latency and high availability. 
 It is widely accepted that programming over eventually consistent data type stores is challenging, since arbitrary divergence among replicas is allowed. 
 Moreover, pragmatic protocols actually achieve consistency guarantees stronger than eventual consistency, which can be and need to be utilized to facilitate the reasoning of and programming over replicated data types.
Toward the challenges above, we propose the \vs framework for precise measurement of eventual consistency semantics.
\vs employs the visibility-arbitration specification methodology in concurrent programming, which extends the linearizability-based specification methodology with a dynamic visibility relation among operations. 
The consistency measurement using \vs is NP-hard in general. To enable practical and efficient consistency measurement in replicated data type stores, the \vs framework refactors the existing brute-force checking algorithm to a generic algorithm skeleton, which further enables efficient pruning of the search space and effective parallelization.
We employ the \vs framework for consistency measurement in two replicated data type stores Riak and CRDT-Redis. 
The experimental evaluation shows the usefulness and cost-effectiveness of \vs in realistic scenarios.    

\end{abstract}

\maketitle

\pagestyle{\vldbpagestyle}
\begingroup\small\noindent\raggedright\textbf{PVLDB Reference Format:}\\
\vldbauthors. \vldbtitle. PVLDB, \vldbvolume(\vldbissue): \vldbpages, \vldbyear.\\
\href{https://doi.org/\vldbdoi}{doi:\vldbdoi}
\endgroup
\begingroup
\renewcommand\thefootnote{}\footnote{\noindent
This work is licensed under the Creative Commons BY-NC-ND 4.0 International License. Visit \url{https://creativecommons.org/licenses/by-nc-nd/4.0/} to view a copy of this license. For any use beyond those covered by this license, obtain permission by emailing \href{mailto:info@vldb.org}{info@vldb.org}. Copyright is held by the owner/author(s). Publication rights licensed to the VLDB Endowment. \\
\raggedright Proceedings of the VLDB Endowment, Vol. \vldbvolume, No. \vldbissue\ %
ISSN 2150-8097. \\
\href{https://doi.org/\vldbdoi}{doi:\vldbdoi} \\
}\addtocounter{footnote}{-1}\endgroup

\ifdefempty{\vldbavailabilityurl}{}{
\vspace{.3cm}
\begingroup\small\noindent\raggedright\textbf{PVLDB Artifact Availability:}\\
The source code, data, and/or other artifacts are available at \url{\vldbavailabilityurl}.
\endgroup
}

\begin{CJK*}{UTF8}{gbsn}

\section{Introduction}


Large-scale distributed systems often resort to replication techniques to achieve fault-tolerance and load distribution \cite{Burckhardt14, Shapiro11a, Enes19}. For a large class of applications, user-perceived latency and overall service availability are widely regarded as the most critical factors. Thus, many distributed systems are designed for low latency and high availability in the first place \cite{Lloyd13}. These systems have to trade data consistency for high availability and low latency \cite{Brewer12, Gilbert12}. A common approach is to resort to \textit{eventual consistency}, i.e., allowing data replicas to temporarily diverge, and making sure that these replicas will eventually converge to the same state in a deterministic way. The Conflict-free Replicated Data Type (CRDT) framework provides a principled approach to maintaining eventual consistency \cite{Shapiro11a, Burckhardt14}.

CRDTs are key components in modern geo-replicated systems, such as Riak \cite{Riak}, Redis-Enterprise \cite{Redis-Enterprise}, and Cosmos DB \cite{CosmosDB}. Though CRDT designs and implementations widely use the notion of eventual consistency, this is a rather fuzzy consistency term and covers a broad range of actual consistency behaviors. Providing only eventual consistency makes programming over CRDTs difficult, in the sense that there can be arbitrary divergence between the replicas \cite{Bermbach14, Lloyd13}. 

The tension between the wide use of eventual consistency and the difficulty of programming over eventually consistent CRDTs motivates the measurement of eventual consistency \cite{Bailis14-vldbj, Burckhardt14-book, Golab14}. On the one hand, both providers and consumers of the distributed storage service benefit from knowing the precise degree of data (in)consistency. Quantifying the degree of data inconsistency can even enable consumers and service providers to negotiate compensations proportional to the severity of violations in certain scenarios \cite{Golab11}.
On the other hand, pragmatic protocols, e.g. in Cassandra, Mongodb, and Eiger, actually achieve consistency guarantees stronger than eventual consistency \cite{Lloyd13}. 
Such stronger consistency semantics can be and should be utilized. Precise reasoning about the consistency semantics will significantly ease the burden of programming over CRDT data stores.

Existing consistency measurement schemes are mainly based on certain metrics delineating the staleness of data, e.g., how long the data has been stale \cite{Golab14} and how many new versions of data have been generated \cite{Golab18}.
Since the number of time units and that of data versions are numerical values, they simplify the issue of comparing different consistency semantics.
However, these consistency measurement schemes are external, i.e., they only describe in a black-box manner how consistent the data observed is.
We argue that, in order to better facilitate the programming over CRDTs, we need  white-box consistency measurement schemes which can give insights into how the data updates are propagated and observed across data replicas.

In this work we investigate a methodology for the precise measurement of eventual consistency semantics for CRDTs.
We employ a formal specification framework for concurrent objects \cite{Bermbach14, Emmi18, Emmi19}, which is based on the \textit{visibility} and \textit{arbitration} relation over operations in a data type access trace.
We name it the \textit{Visibility-Arbitration} framework (\visar in short).
The \visar framework extends the linearizability-based specification methodology with a dynamic visibility relation among operations, in addition to the standard dynamic happen-before and linearization relations. 
The arbitration relation is a total order and indicates how the system resolves conflicts among concurrent events. 
The visibility relation is an acyclic relation that describes the relative timing of update propagation among operations. 
The weaker the visibility relation, the more an operation is allowed to ignore the effects of concurrent operations that are linearized before it. 
We use diverse constraints on the visibility relation to define different levels of data consistency.

Existing consistency measurement schemes based on the \visar framework are mainly designed for concurrent objects, 
which are optimized for high-performance multi-threaded code \cite{Emmi19, Emmi18}. 
Since concurrent objects often employ lock-free implementations and do not provide atomicity, the \visar framework is designed to measure the deviation from atomicity. 
However, in our scenarios, the replicated data types are designed to provide eventual consistency in the first place. 
Our main objective is not to measure the deviation from strong consistency, but to measure the augmentation over weak consistency.

Though consistency measurement using the \visar framework is NP-hard in general \cite{Gibbons97, Emmi18}, brute-force checking algorithms are employed in practice \cite{Emmi19}. 
The key insight to circumvent the NP-hardness obstacle is that the violations are often local. That is, only a few operations can witness the violation. 
The small workload is repeated a sufficient number of (often millions of) times to find and measure the deviation from atomicity. 
However, the problem of eventual consistency measurement does not have the locality existing brute-force checking algorithms depend on.
We are confronted with the challenge of further optimizing existing checking algorithms.

Our intuition is that semantic knowledge of the data type under checking can be leveraged to reduce the measurement cost, while moderately sacrificing the generality of the measurement scheme.
Following this intuition, our first challenge is to refactor the brute-force consistency checking algorithm to a generic algorithm skeleton, which can conveniently incorporate semantic-aware optimizations.
The second challenge is to find the semantic information which can effectively reduce the checking cost while ensuring correct consistency measurement.

Toward the challenges above, we transform the brute-force checking algorithm for consistency measurement to a non-recursive algorithm skeleton.
Note that, the basic rationale of the measurement based on \visar is to enumerate all possible permutations of operations as well as all possible visibility relations among operations.
In the non-recursive algorithm skeleton, the intermediate state of enumeration is explicitly maintained at runtime.

Semantic knowledge of the data type under checking is expressed as \textit{pruning predicates}.
As indicated by its name, the pruning predicate prunes the state space while ensuring correct consistency checking.
We extract the pruning predicates from subhistories pertaining to one single data element. 
We enumerate all valid abstract executions over the subhistory and extract ``facts" which hold for all possible abstract executions.
Such facts are expressed as pruning predicates.
Given the generic algorithm skeleton for consistency checking, the pruning predicates can be flexibly ``plugged" into the \vs skeleton.
Whenever \vs finds violation of the pruning predicates, it can safely terminate the current search.
Moreover, since the construction of different abstract executions are independent, the search can be parallelized in a straightforward way.
We also periodically balance the load among the parallel worker threads.

We conduct consistency measurement experiments with \vs on two replicated data type stores Riak \cite{Riak} and CRDT-Redis \cite{CRDT-Redis}. 
The measurement quantifies the differences between different conflict resolution strategies, as well as the differences between different configurations of the replicated data type store. The consistency measurement results are useful for both replicated data type designers and upper-layer application developers.
The experiments also show the effects of our two optimizations and demonstrate the cost-effectiveness of \vs.

The rest of this work is organized as follows.
Section \ref{Sec: Preli} introduces the preliminaries of CRDTs and the \visar framework.
Section \ref{Sec: VS} presents the \vs framework, and Section \ref{Sec: Exp} presents the experimental evaluation.
Section \ref{Sec: RW} discusses the related work. 
In Section \ref{Sec: Concl}, we conclude this work and discuss the future work.

\section{Preliminaries} \label{Sec: Preli}

In this section, we first introduce the basic concepts of Conflict-free Replicated Data Types (CRDTs). Then we introduce the Visibility-Arbitration (\textsc{Vis-Ar}) framework and the consistency measurement based on visibility relaxation.

\subsection{Replicated Data Type}

We basically adopt the system model in \cite{Jiang20}.
We consider a replicated data store storing \textit{objects} of some abstract data type.

\begin{definition}
    (Abstract Data Type). An abstract data type $\tau \in \textit{Type}$ is a pair $\tau = (\textit{Op}, \textit{Val})$ such that:
    \begin{itemize}
        \item \textit{Op} is the set of operations supported by $\tau$;
        \item \textit{Val} is the set of values allowed by $\tau$. We use $\perp \in \textit{Val}$ to denote that there is no return value for some operation. 
    \end{itemize}
\end{definition}

\begin{definition}
    (Sequential Semantics). The sequential semantics of a data type $\tau \in \textit{Type}$ is defined by a function $\mathcal{F}_{\tau}: \textit{Op}^{*} \times \textit{Op} \rightarrow \textit{Val}$. It means that given a sequence of operations $S$ and an operation $o$, the return value of $o$ is $\mathcal{F}_{\tau}(S, o) \in \textit{Val}$ when $o$ is performed right after $S$. 
\end{definition}

We use ``$m(arg_1, \ldots, arg_n)\Rightarrow r$" to denote the invocation of a method $m$, with arguments $arg_1, \ldots, arg_n$, resulting in a return value $r$. 
We use the sequential semantics to explain the execution of operations. The return value of an operation is determined by a certain sequence of operations that are performed before or concurrent with it. 

In this work, we consider operations which are either \textit{queries} or \textit{updates}. 
Query operations only inspect state of the data type and do not change the state, while update operations modify state of the data type. 
Some data types may provide query-update primitives. 
For example, the $extract\text{-}max()$ primitive of a priority queue is a query-update primitive. 
We can 
decompose each query-update operation into a pair of query and update \cite{Wang19}. 
So we only consider query operations and update operations in this work. 



\subsection{\visar Framework}

A consistency model acts as the contract between the underlying data type store and the upper-layer applications.
The \visar framework is an axiomatic consistency specification of concurrent objects \cite{Bermbach14, Emmi18, Emmi19}. 
It extends the linearizability-based specification methodology with a dynamic visibility relation among operations, in addition to the standard dynamic happens-before and linearization relations. 
The weaker the visibility relation, the more an operation is allowed to ignore the effects of concurrent operations that are linearized before it. 
Tuning the definition of the visibility relation gives us different levels of data consistency. We use various predicates (i.e., constraints) on the visibility relations to define different levels of data consistency.

Histories, i.e., traces of data type access, are observable behaviors of a data store. 
The history defines how the clients interact with the data type store, and specifies what information is recorded about this interaction.
\begin{definition}
    (History). A history is an event graph represented by a tuple $\mathcal{H} = (E, op, rval, so)$ such that:
    \begin{itemize}
        \item $E$ is the set of all events of operations performed by clients in a single execution; 
        \item $op : E \rightarrow Op$ describes the operation of an event; 
        \item $rval : E \rightarrow Val$ describes the return value of the operation $op(e)$ of an event $e$; 
        \item $so \subseteq E \times E$ is a partial order over $E$, called \textit{session order}. It describes the order of operations invoked by clients within a session. 
    \end{itemize}
\end{definition}

In this work, we only record the order between operations within the same session. The \textit{happen-before} relation 
is limited to events/operations within the same session. 
We do not record the \textit{return-before} order over events which captures the real-time ordering of non-overlaping operations \cite{Burckhardt14-book}, since we do not have a global time in distributed scenarios. 


In order to tell whether a history is consistent or not, the basic rationale underlying the \visar framework is that, a history is consistent if and only if we can justify it, by augmenting it with some additional information that explains the observed return values. 

Based on the \visar framework, we add arbitration (i.e., linearization) and visibility relations to a history, and the augmented history is called an \textit{abstract execution}.
\begin{definition}
    (Abstract Execution). An abstract execution is a triple $\mathcal{A} = ((E, op, rval, so), vis, ar)$ such that:
    \begin{itemize}
        \item $(E, op, rval, so)$ is a history;
        \item Visibility relation $vis \subseteq E \times E$ is an acyclic and natural relation. A relation $R \subseteq A \times A$ is natural if $\forall x \in A : \left\vert R^{-1}(x) \right\vert < \infty$;
        \item Arbitration relation $ar$ is a total order on $E$. 
    \end{itemize}
\end{definition}

If an operation $a$ is visible to $b$ (denoted by $a \stackrel{vis}{\rightarrow} b$), it means that the effect of $a$ is visible to the client performing $b$. 
If an operation $a$ is arbitrated before $b$ (denoted by $a \stackrel{ar}{\rightarrow} b$), it means that the system considers operation $a$ to happen earlier than operation $b$. 
We use $lin$ to denote the linearization of operations after we decide the arbitration for concurrent operations. 

\setlength\emergencystretch{\hsize}


\begin{definition}
    (Consistency Model). A consistency model is a set of consistency guarantees on abstract executions. 
    A consistency guarantee $\mathcal{\phi}$ is a constraint or property of an abstract execution $\mathcal{A}$. 
\end{definition}

We write $\mathcal{A} \models \mathcal{\phi}$ if the abstract execution $\mathcal{A}$ satisfies the consistency guarantee $\mathcal{\phi}$. 
We can define a large variety of consistency guarantees by formulating conditions on the various attributes and relations appearing in the abstract execution.
An abstract execution $\mathcal{A}$ satisfies consistency model $\Phi = \{\mathcal{\phi}_1, \mathcal{\phi}_2, \ldots, \mathcal{\phi}_n\}$, if it satisfies every consistency guarantees in $\Phi$. 
%
A history $\mathcal{H}$ satisfies the consistency model $\Phi$, denoted $\mathcal{H} \models \Phi$, if it can be extended (by adding visibility and arbitration) to an abstract execution that satisfies consistency model $\Phi$. Formally, $\mathcal{H} (E, op, rval, so) \models \Phi \Leftrightarrow \exists vis, ar : ((E, op, rval, so), vis, ar) \models \Phi$. 

There are mainly two types of consistency guarantees. 
One is about the ordering of operations. This type of guarantees mainly consists of constraints on arbitration and visibility relations.
The other is about the data type semantics that gives meanings to the operations. 

\subsection{Consistency Measurement via Visibility Relaxation} \label{SubSec: Vis-Relax}

In \cite{Emmi19}, a consistency spectrum consisting of six consistency levels is defined by visibility relaxation. 
These consistency levels, from weak to strong, are \textit{weak}, \textit{basic}, \textit{monotonic}, \textit{peer}, \textit{causal} and \textit{complete} visibility.
The two ends of the consistency spectrum are weak and complete visibility. The weak visibility permits any observation made by an operation, while the complete visibility requires that operations observe all operations which are arbitrated before them. 

Between these two ends, the basic visibility requires operations to see their happen-before predecessors which corresponds to a strengthening of the read-your-writes guarantee. 
The monotonic visibility requires each operation to see those operations its happen-before predecessors see. 
It corresponds to the monotonic reads guarantee. 
On the basis of the monotonic visibility, the peer visibility requires a read to observe an update only if it has observed all the predecessors from the update's session.
This corresponds to the monotonic writes guarantee.
The causal visibility requires that the visibility relation be transitive. It means that each operation is required to see those operations visible to the operations it sees. 

We define a set of predicates which constraint the observation made by a given operation. 
Formally, a visibility predicate $\varphi(o, \mathcal{A})$ is a first-order predicate on operation $o$ in an abstract execution $\mathcal{A}$. Given an abstract execution $\mathcal{A}$, we define $vis(o)$ is a set of operations visible to $o$; $hb(o)$ is a set of operations that happens before $o$ according to the session order $so$; $lin(o)$ is a set of operations arbitrated before $o$. 
\begin{itemize} \label{vis-level}
    \item weak$(o, \mathcal{A}) \Leftrightarrow \textbf{\textit{true}}$
    \item basic$(o, \mathcal{A}) \Leftrightarrow hb(o) \subseteq vis(o)$
    \item monotonic$(o, \mathcal{A}) \Leftrightarrow \forall o' \in hb(o): vis(o') \subseteq vis(o)$
    \item peer$(o, \mathcal{A}) \Leftrightarrow$ monotonic$(o, \mathcal{A}) \land \forall o' \in vis(o): hb(o') \subseteq vis(o)$ 
    \item causal$(o, \mathcal{A}) \Leftrightarrow$ basic$(o, \mathcal{A}) \land \forall o' \in vis(o): vis(o') \subseteq vis(o)$ 
    \item complete$(o, \mathcal{A}) \Leftrightarrow lin(o) \subseteq vis(o)$
\end{itemize}

\begin{figure}[tb]
    \includegraphics[width=\linewidth]{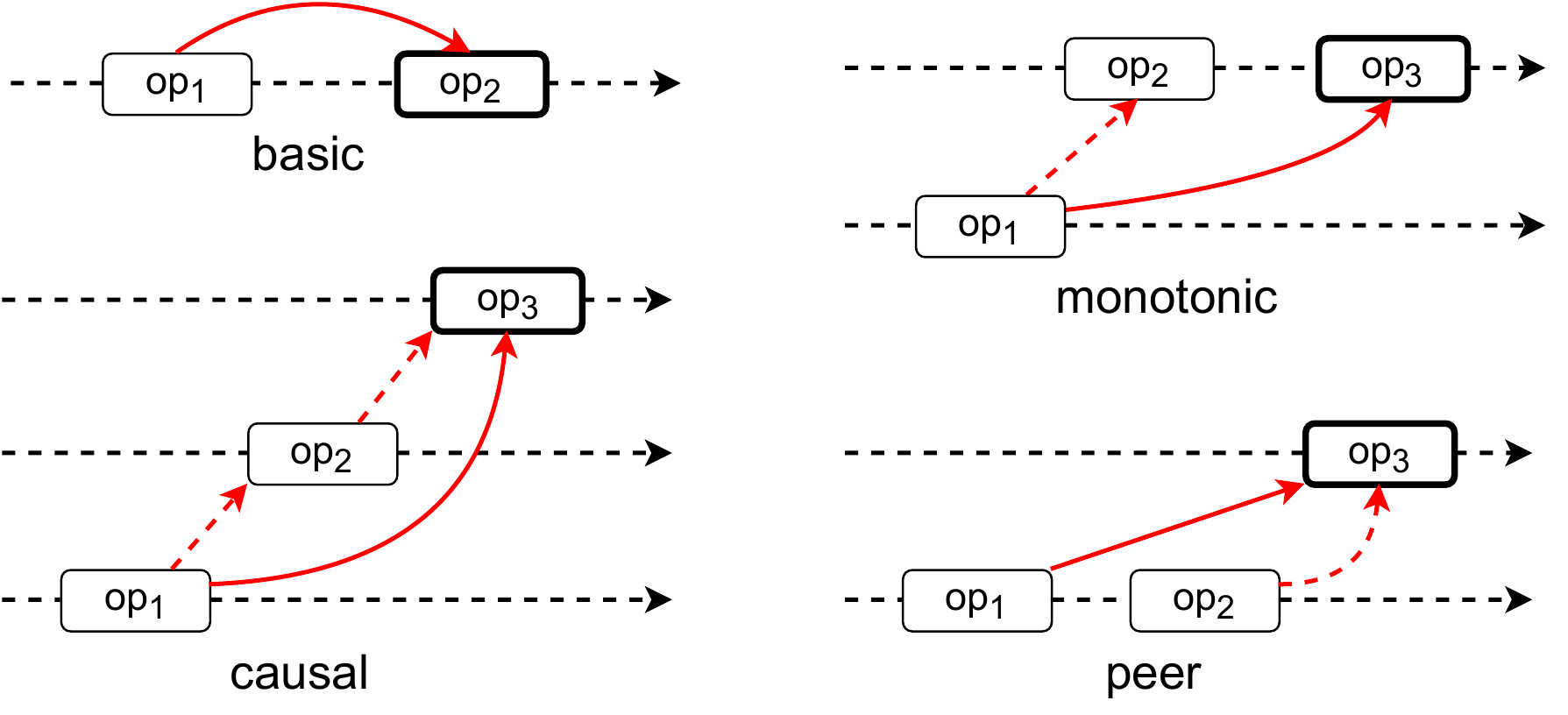}
    \centering
    \caption{Examples of four consistency levels.}
    \label{F: Visibility}
\end{figure}

In Fig.~\ref{F: Visibility} we illustrate these four consistency levels by examples, which are inspired from \cite{Emmi19}. 
The operations (rounded rectangle) visible to the given operation (bold rounded rectangle) depend on the visibility level. 
The \textit{basic} visibility implies visibility (solid arrow) of happens-before predecessors (horizontal line) while the \textit{monotonic} visibility implies visibility of those visible (dashed arrow) to predecessors. 
The \textit{peer} visibility implies visibility of predecessors of visible operations, and the \textit{causal} visibility implies transitive visibility. 
In the rest of this work, we use this 6-level scale of consistency levels for the consistency measurement.



Basically, the consistency measurement is conducted on one history or trace of replicated data type access. Theoretically speaking, to measure the consistency level of one data type, we need to measure all possible (infinitely many) traces of this data type, which is simply not possible. Moreover, it is usually quite difficult to strictly prove the consistency level certain replicated data type provides. In this work, when we say consistency measurement for one data type, we mean that a sufficiently large number of traces generated by this data type are measured, and the result of this measurement is a sufficiently accurate approximation of the consistency level provided by this data type (more details can be found in Section \ref{SubSec: Exp-Design}).

\section{\vs Framework} \label{Sec: VS}

In this section, we first overview the basic design rationale behind the \vs framework. Then we present the key techniques. In the next Section \ref{Sec: Exp}, we will demonstrate how to apply \vs in realistic data type store scenarios.

\subsection{\vs  Overview}

Consistency measurement using \vs boils down to the search of an abstract execution satisfying certain constraints on the arbitration and visibility relations. 
This search process involves enumerating the linearizations, i.e. arbitrations of all operations, as well as the visibility relations among operations.
To prove that one trace satisfies certain consistency model, it is sufficient to find one single abstract execution that satisfies all the constraints. 
We name this abstract execution the \textit{certificate abstract execution} (there could be more than one certificate execution).

The search process is intractable in general \cite{Gibbons97, Emmi19, Emmi18}. However, the search can still be practical and cost-effective in replicated data type store scenarios. 
To enable efficient search of abstract executions, we first refactor the existing recursive search algorithm to a non-recursive algorithm skeleton.
The intermediate states for construction of the abstract executions are explicitly maintained by the algorithm skeleton at runtime.

Based on the \vs skeleton, we propose optimizations from two orthogonal dimensions. 
First, the non-recursive algorithm skeleton enables efficient pruning of the search space. 
We can flexibly plug pruning predicates into the algorithm skeleton to avoid the unnecessary search of certain abstract executions. 
The pruning predicates can be automatically extracted from the subhistory concerning multiple update operations and one query operation on one single data element. 
Second, we find that the enumeration of different candidate executions are independent. Thus we further parallelize the search. 
The non-recursive search skeleton also facilitates the load balancing among the parallel worker threads.

\subsection{\vs Algorithm Skeleton}

We first briefly review the existing recursive search algorithm \cite{Emmi19, Emmi18}. Then we present the non-recursive algorithm skeleton.

\subsubsection{The Original Recursive Search Algorithm}

Consistency measurement using the \vs framework can simply use a recursive backtracking algorithm to search the possible arbitration and visibility relations among operations in a brute-force manner. We use $\ang{lin, vis}$ to denote an abstract execution. If an abstract execution $\ang{lin, vis}$ satisfies a consistency model $\Phi$, we have $\langle lin, vis \rangle \models \Phi$. 
During the search process, we manipulate some prefix of $lin$, denoted by $lin_i$. We use $vis_i$ to denote the visibility relation on operations in $lin_i$. 
The search algorithm can be viewed as extending candidate executions from $\ang{lin_i, vis_i}$ to $\ang{lin_{i+1}, vis_{i+1}}$.

\setlength\emergencystretch{\hsize}

The search algorithm tries all possible extensions from the current ``candidate execution" $\ang{lin_i, vis_i}$. 
When valid extensions cannot be obtained, the current candidate is discarded and the search algorithm backtracks to another candidate execution.
When we finally extend the prefix to a whole abstract execution containing all the operations in the data type access trace, we successfully get the certificate abstract execution, and the search algorithm returns \textsf{TRUE}.
If no certificate execution can be obtained after trying all possible candidates, the search algorithm returns \textsf{FALSE}.

\subsubsection{The Refactored Non-recursive Search Skeleton} \label{SubSubSec: Revised-VS}

The main improvement made to the recursive search algorithm is that we refactor it to a non-recursive algorithm skeleton. 
This means that, the algorithm skeleton itself maintains a double-ended queue (\textit{deque} in short) to store the intermediate search states $\ang{lin_i, vis_i}$ at runtime.
The explicit maintenance of intermediate search states greatly facilitates further pruning of unnecessary searches.

\begin{figure}[tb]
    \includegraphics[width=\linewidth]{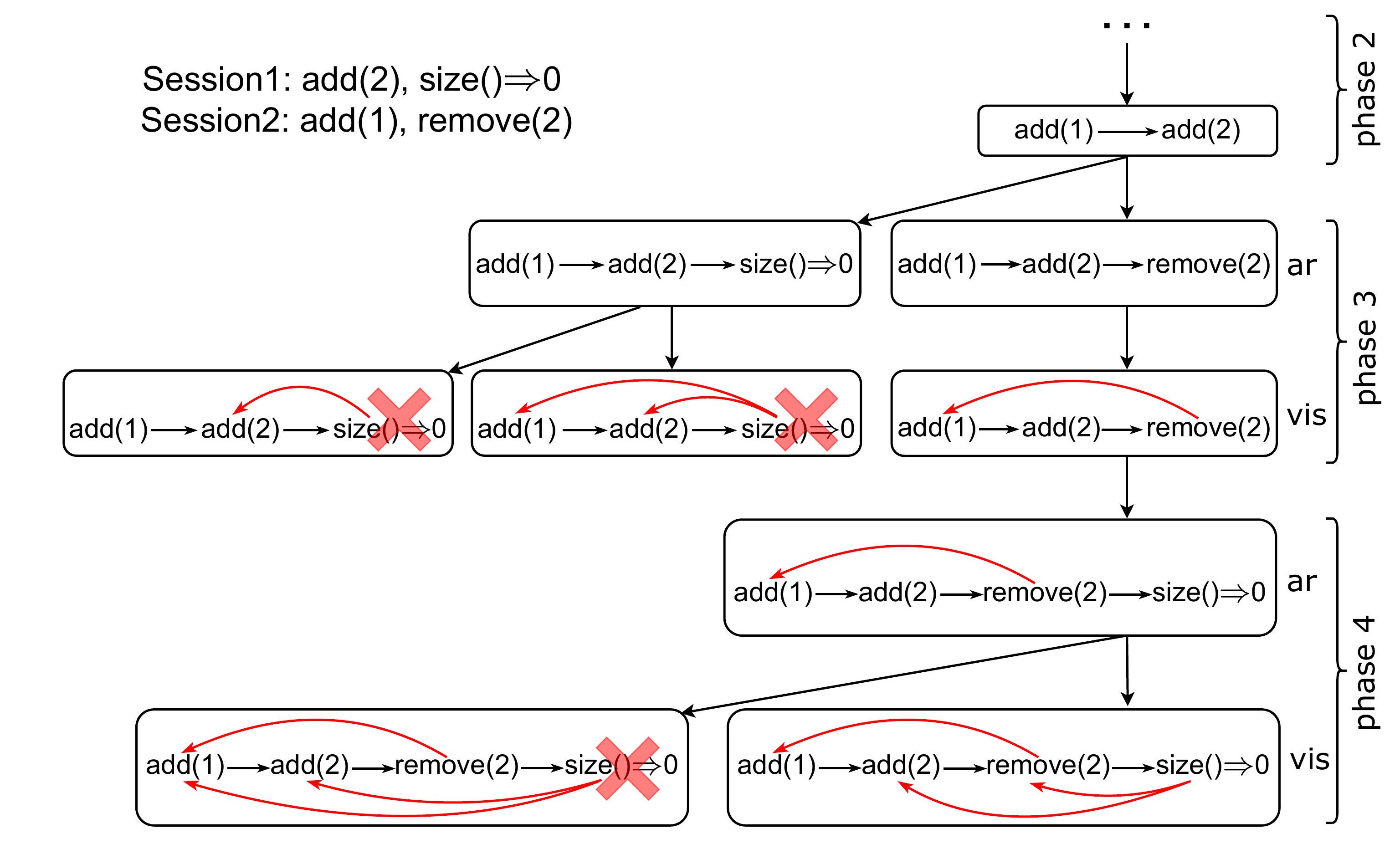}
    \centering
    \caption{Search of the certificate execution.}
    \label{F: Tree}
\end{figure}

The non-recursive \vs algorithm searches abstract executions through phases, as shown in Fig.~\ref{F: Tree}. 
In each phase, it extends $\ang{lin_i, vis_i}$ to $\ang{lin_{i+1}, vis_{i+1}}$. 
There are two steps in each phase. 
In the ``ar step", we add one operation to $lin_i$ and obtain $lin_{i+1}$. 
In the ``vis step", we extend $vis_i$ to $vis_{i+1}$ by adding visibility relations concerning the newly added operation. 

As shown in Fig.~\ref{F: Tree}, the algorithm has already constructed a certificate abstract execution $\ang{lin_2, vis_2}$ that contains operations \textit{add}(1) and \textit{add}(2). Then the algotirhm starts phase 3. 
In the ar step, the algorithm first appends operation \textit{size}() to the $lin_2$ and obatins $lin_3$. 
In this example, we require each operation be visible to operations within the same session. 
So the operation \textit{size}() can return 1 if $add(1) \stackrel{vis}{\rightarrow} size()$, or return 2 if $\{add(1), add(2)\} \stackrel{vis}{\rightarrow} size()$. 
But neither of these two visibility relations bring the required return value $size \Rightarrow 0$. 
Thus the algorithm backtracks for other candidates.

Then we append operation \textit{remove}(2) to the $lin_2$. We can construct a certificate abstract execution $\ang{lin_3, vis_3}$ and finish phase 3. 
Then we proceed to phase 4, and the algorithm appends the last operation $size\Rightarrow0$ to the linearization. 
There are four possible cases of visibility extension concerning the newly added operation $size\Rightarrow0$: 
$\left\{ add(2) \right\}\stackrel{vis}{\rightarrow} size()\Rightarrow 0$, 
$\left\{ add(1), add(2) \right\}\stackrel{vis}{\rightarrow} size()\Rightarrow 0$,
$\left\{ add(2), remove(2) \right\}\stackrel{vis}{\rightarrow} size() \Rightarrow 0$, 
and $\left\{\textit{add(1)}, \textit{add(2)}, \textit{remove}(2)\right\}\stackrel{vis}{\rightarrow}\textit{size}()$.
We only show two of the four possible cases in Fig.~\ref{F: Tree}, due to the limit of space. 
The operation \textit{size}() returns 0 only if $\{add(2), remove(2)\} \stackrel{vis}{\rightarrow} size()$. 
Finally, we find a certificate abstract execution, and prove that the history satisfies the \textit{basic} consistency level.

Upon initialization, we build a \textit{search state deque} with the initial search state $\ang{lin_0, vis_0}$. State $\ang{lin_0, vis_0}$ contains no operation and has an empty visibility relation. 
During the search process, we keep polling search states from the head of the deque.
The newly popped search state is verified by the method \textsf{isValidAbstractExecution} ({Line 6 in Algorithm \ref{A: VS}}). This method basically checks whether the return values of all query operations in the current search state are allowed by the sequential semantics \textit{Impl} of the data type.
Invalid search states are discarded and valid ones are extended as follows. 

For a given valid search state $\ang{lin_i, vis_i}$, we first extend the linearization by \textsf{linExtend} (Line 10 in Algorithm \ref{A: VS}), and get a set of new search states:
$$newLin = \left\{ \ang{lin_{i+1}^1, vis_i}, 
\ang{lin_{i+1}^2, vis_i}, \ldots, \ang{lin_{i+1}^p, vis_i} \right\}.$$

\noindent Then we extend visibility relations according to the newly added operation.
Each newly-extended \textit{search state} $\ang{lin_{i+1}^{r}, vis_i}$ will be extended to a set of \textit{search states}:
\begin{eqnarray*}
    newVis_r &=& \{ \ang{lin_{i+1}^{r}, vis_{i+1}^1}, \ang{lin_{i+1}^{r}, vis_{i+1}^2}, \ldots, \\
    &\;& \ang{lin_{i+1}^{r}, vis_{i+1}^{q}} \}
\end{eqnarray*}

\noindent Each search state in $\bigcup_{i=1}^{p} newVis_i$ is a possible extension of $\ang{lin_i, vis_i}$.We push all these candidate states into the head of the deque. 
In the next step, the search algorithm will get a candidate from the head of the deque and repeat this extension process. 
Once the search state cannot be extended any more, the algorithm will backtrack by polling older candidates from the head of the deque. 

The {\vs} algorithm returns \textsf{TRUE} if it obtains a certificate abstract execution.
The algorithm returns \textsf{FALSE} if it runs out of candidates but still cannot obtain any certificate execution.
The pseudocode of the \vs algorithm skeleton is listed in Algorithm \ref{A: VS}.

\begin{algorithm}[tb]
    \caption{The \vs Algorithm Skeleton}
    \label{A: VS}
    \KwIn{history $\mathcal{H}$, 
        data type implementation \textit{Impl}, 
        consistency model $\Phi$}
    \KwOut{whether history $\mathcal{H}$ satisfies the given consistency model $\Phi$}
    \textit{deque} $\leftarrow$ \{\ \}\\
    \textit{initState} $\leftarrow$ $\ang{\emptyset, \emptyset}$\\
    \textit{deque}.\textsf{push}(\textit{initState})\\
    \While{$\lnot$deque.\textnormal{\textsf{isEmpty}( )}}{
        \textit{state} $\leftarrow$ \textit{deque}.\textsf{pollHead}( )\\
        \If{\textnormal{\textsf{isValidAbstractExecution}(}Impl, state\textnormal{)}}{
            \If{state\textnormal{.\textsf{isComplete}( )}}{
                \Return{\textnormal{\textbf{true}}}
            }
            \textit{newLin} $\leftarrow$ \textit{state}.\textsf{linExtend}($\mathcal{H}$)\\
            \ForEach{\textit{newLinState} $\in$ \textit{newLin}}{
                \textit{newVis} $\leftarrow$ \textit{newLinState}.\textsf{visExtend}($\Phi$)\\
                \text{$\langle$search space pruning plugged here$\rangle$}\\
                \textit{deque}.\textsf{pushHeadAll}(\textit{newVis})\\
            }
        }
    }
    \Return{\textnormal{\textbf{false}}}
\end{algorithm}

\subsection{Pruning of the Search Space}

The key to circumventing the NP-hardness obstacle of abstract execution search is to leverage the semantic knowledge of the data type under search and prune the search space. 
Thus, we inspect the data type access trace and automatically extract \textit{pruning predicates}, which can be plugged into the \vs algorithm skeleton (Algorithm \ref{A: VS}). 
For each candidate abstract execution, the search algorithm checks the pruning predicates. 
Once any pruning predicate is detected \textsf{FALSE}, the candidate abstract execution is directly discarded.
We will prove that the pruning process is ``safe", in the sense that the pruning process never discards any candidate abstract execution which can be extended to a certificate abstract execution.

In this section, we first present the basic rationale of the pruning algorithm with three illustrative examples.
Then we present the design and correctness proof of the pruning algorithm.

\subsubsection{Rationale of Pruning}

The search algorithm enumerates every valid abstract execution in a brute-force way.
The insight behind our pruning scheme is that, for a  given data element in a container type, the update and query operations concerning this element cannot be linearized in an arbitrary way. Moreover, the visibility relations among can only have limited patterns.

For example, for each read operation \textit{r}, there exists a unique write operation \textit{w}. 
The operation \textit{w} must happen before \textit{r}. 
The rule mentioned above can be formally defined as the write-to order \cite{Wei16}. 
However, the \vs algorithm is ignorant of this type of knowledge concerning one specific data element, and enumerates all abstract executions it thinks that is possible.
Thus, we can extract this type of knowledge and express it as pruning predicates. 
We will use these pruning predicates before adding new search states into the deque ({Line 13 in Algorithm \ref{A: VS}}). 

Based on the observation above, we introduce the notion of \textit{query cluster}. 
All operations in the query cluster are associated with the same data element.
The query cluster contains exactly one query operation, which is denoted the \textit{dictated query}. The query cluster contains one or more update operations, which are denoted the \textit{dictating updates}. The dictating updates are those update operations which can potentially affect the return value of the dictated query. 

We consider two situations in extraction of the pruning predicates. One situation is that the update operations are the necessary and sufficient condition for the result of a query operation. For example, whether the set contains the element $x$ is determined by the arbitration and visibility of all $add$ and $remove$ operations associated with element $x$. 
The other situation is that the update operations are only the necessary condition for the result of a query operation. For example, $get\_max$ returns that the maximum element of the priority queue is $x$ with value $v$. All update operations associated with element $x$ only guarantee that there exists an element $x$ with value $v$ in the priority queue at that point.
But only these update operations cannot decide the return value of $get\_max()$.
Fortunately, the necessary condition is enough to ensure the safety of the pruning algorithm. 
In extraction of the pruning predicates, we do not consider operations which correspond to multiple data elements, e.g., the $isEmpty()$ operation, which returns whether the data container is empty.




%
%
%
%
%
%

We will use two data types \textit{set} and \textit{priority queue} as illustrative examples. 
The set is a container of elements, in which each element is identified by its unique \textit{id}. 
The following primitives are involved in our examples: 
\begin{itemize}
    \item $add(e)$ : add element $e$. 
    \item $remove(e)$ : remove element $e$. 
    \item $contains(e)$ : return \textsf{TRUE} if $e$ is in the set.
\end{itemize}

\noindent The priority queue is a container of elements of the form $e = (id, priority)$.
The following primitives are involved in the examples: 
\begin{itemize}
    \item $insert(e, x)$ : enqueue element $e$ with initial priority $x$. 
    \item $inc(e, \delta)$ : increase the priority of element $e$ by $\delta$ (may be negative). 
    \item $get\_pri(e)$ : return the priority value of $e$.
    \item $get\_max()$ : return the \textit{id} and \textit{priority} of the element with the highest priority. 
\end{itemize}


%

The pruning predicate is obtained by inspecting the query cluster pertaining to one specific data element.
There can be two types of pruning predicates. 
The first type is simpler, which only considers the arbitration relation between operations, while the second type considers both the arbitration and the visibility relations.


By in-depth inspection of the data type access trace and manual exploration of the brute-force checking process, we come up with three predicate templates, i.e. patterns of pruning predicates we want to extract.
The first predicate template focuses on the arbitration relation. It aims to find that two operations always have the arbitration relation, no matter how you construct the linearization of operations. Specifically, $\tarb$ states that operation $x$ must be arbitrated before operation $y$ in every search state $\ang{lin, vis}$:
\begin{eqnarray*}
    \tarb(x, y) \mydef \forall \ang{lin, vis}: (x,y) \in lin
\end{eqnarray*}

\noindent To extract pruning predicates following the template $\tarb$, we enumerate all valid abstract executions of the subhistory consisting of all operations in one query cluster.

\begin{figure}[tb]
    \includegraphics[width=\linewidth]{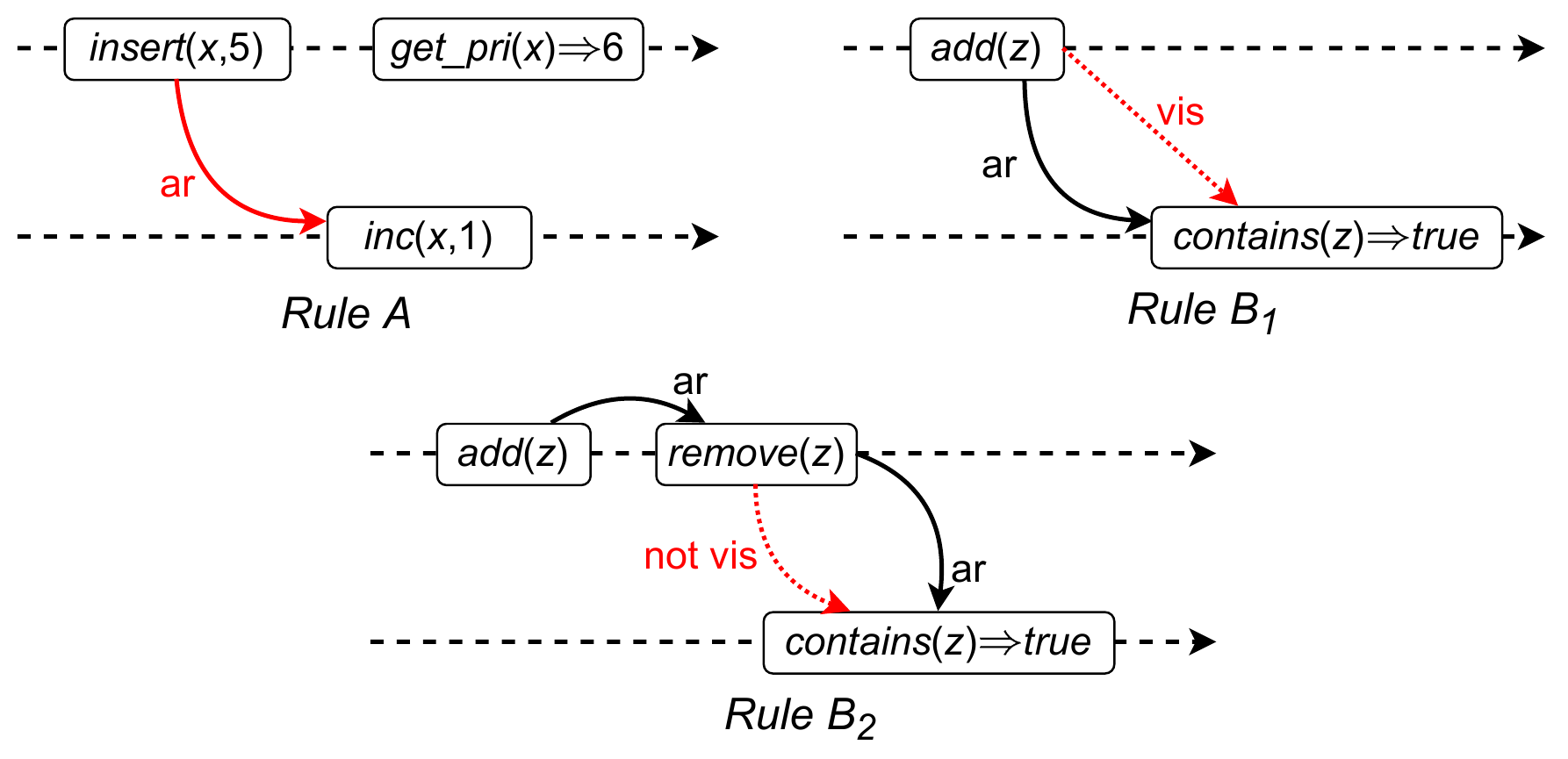}
    \centering
    \caption{Examples of three predicate templates.}
    \label{F: Template}
\end{figure}

For example, assume that there is a subhistory with two client sessions, as shown in Fig.~\ref{F: Template}.
One session has operation $insert(x,5)$, followed by $inc(x,1)$.
The second session has one single operation $get\_pri(x) \Rightarrow 6$.
For ease of illustration, we simply assume that there are no other operations on $x$ in the whole trace. 
It is evident that the return value of $get\_pri(x)$ can be 6 only if it is linearized after $inc(x,1)$.
All valid abstract executions of this query cluster must have this arbitration order $inc(x,1)  \stackrel{ar}{\rightarrow} get\_pri(x) \Rightarrow 6$.
This fact must also hold for all candidate abstract executions the \vs algorithm constructs. Thus it can be extracted as a pruning predicate and plugged into the \vs algorithm skeleton to avoid unnecessary searches.

There are two more predicate templates 
of the second type, which concern both the arbitration and the visibility relations. 
As for the $\tvis$ example in Fig.~\ref{F: Template}, if the arbitration relations among operations have been decided, we can infer that there must be the visibility relation between $remove(z)$ and $contains(z) \Rightarrow \textsf{FALSE}$. Otherwise, the semantics of the set type are violated.
The $\tnotvis$ template is the dual of $\tvis$. If the return value of $contain(z)$ is \textsf{TRUE}, there should be no visibility relation between $remove(z)$ and $contain(z)$.






Generalizing the examples above, $\tvis$ states that certain arbitration relations and certain visibility relation always co-exist:
\begin{eqnarray*}
    \tvis(S_{lin}, x, y) &\mydef& \forall \ang{lin, vis} : \\
    &\ & S_{lin} \subseteq lin \rightarrow (x,y) \in vis
\end{eqnarray*}

\noindent where $S_{lin}$ is a subset of the arbitration relation (a subset of operation pairs having the arbitration relation).
The dual template $\tnotvis$ states that if there are certain arbitration relations, certain visibility relation cannot occur:
\begin{eqnarray*}
    \tnotvis(S_{lin}, x, y) &\mydef& \forall \ang{lin, vis} : \\
    &\ & S_{lin} \subseteq lin \rightarrow (x,y) \not \in vis
\end{eqnarray*}

\noindent We scan all valid abstract executions of the query cluster and check whether there are operation patterns that can be matched by the two predicate templates $\tvis$ and $\tnotvis$.








\subsubsection{Extraction of Pruning Predicates}

Extraction of the pruning predicates is similar to the ``string matching" process. 
The predicate template is analogous to the pattern string to be searched, while all valid abstract executions of a query cluster as a whole is analogous to the text string being scanned.
Note that we are interested in predicates which are \textsf{TRUE} for all abstract executions of a given query cluster.
The query cluster can simply be treated like normal traces of data type access, and we can directly use the brute-force \vs algorithm (Algorithm \ref{A: VS}) to search all valid abstract executions. 
We extract concrete pruning predicates following the templates $\tarb$, $\tvis$ and $\tnotvis$ from the abstract executions. 
We assume that the size of the query cluster is bounded and the extraction of predicates takes constant time.
We validate our assumption and study the cost of predicate extraction in Section \ref{SubSec: Eval-Results}.

For an abstract execution $\pi_t$ = $\ang{{lin}^t, {vis}^t}$, we define that $lin(\pi_t)$ = $lin^t$, $vis(\pi_t)$ = $vis^t$. 
We define the not-visible relation of an abstract execution $\pi_t$ by $notvis(\pi_t)$ = $\{(x,y) \mid (x,y) \in lin^t \land (x,y) \notin vis^t\}$. 
It is straightforward to extract pruning predicates of template $\tarb$ from valid abstract executions. 
We just need to intersect every linearization of valid abstract executions, as shown in Algorithm \ref{A: T-ARB}.
Specifically, linearization of an abstract execution $\pi$ is a total order of operations. For each pair of operations, $(x,y) \in lin(\pi)$ means that operation $x$ is arbitrated before operation $y$ in abstract execution $\pi$. 
The intersection of all linearization is the arbitration order that every valid abstract execution must obey. 
Each concrete operation pair $(op1, op2)$ in the final intersection of linearizations can generate a concrete pruning predicate $\tarb(op1, op2)$. 

\begin{algorithm}[tb]
    \caption{Extraction of Pruning Predicates for $\tarb$}
    \label{A: T-ARB}
    
    \KwIn{a set of abstract executions $\mathcal{A} = \{\pi_1, \pi_2 \ldots \pi_m\}$}
    \KwOut{a set of pruning predicates of $\tarb$}
    
    $S_{\text{arb}}$ $\leftarrow$ $U$\quad\CommentSty{/* U is the complete relation which contains all possible pairs of operations. */}\\
    \ForEach{$\pi \in \mathcal{A}$}{
        $S_{\text{arb}}$ $\leftarrow$ $S_{\text{arb}} \bigcap lin(\pi)$
    }
    
    \ForEach{pair $(x,y) \in S_{\text{arb}}$}{
        \textbf{yield} $\tarb(x,y)$
    }
\end{algorithm}

To extract pruning predicates following the template $\tvis$, we union the visibility relation of all valid abstract executions: $\mathcal{V} = \bigcup_{t=1}^{m} vis(\pi_t)$. 
For each pair $(x,y) \in \mathcal{V}$, it means that operation $x$ is visible to operation $y$ in some valid abstract execution. 
Then we try to generate a concrete predicate of template $\tvis$ for every pair $(x,y) \in \mathcal{V}$. 

We first intersect the linerarization of valid abstract executions that contains the pair $(x,y)$. 
Any valid abstract execution whose visibility relation contains pair $(x,y)$ must obey the arbitration order in the intersection $S_{\text{arb}}$. 

Next we need to ensure that there exists no counterexample for this concrete predicate. 
Because we extract this concrete predicate only from valid abstract executions which contain pair $(x,y)$ in their visibility relation. 
We must check whether it is a necessary condition for those valid abstract executions which do not contain pair $(x,y)$ in their visibility relation.
We discard this concrete predicate, if there is a valid abstract execution whose visibility relation does not contain pair $(x,y)$ but it obeys all arbitration order in the intersection $S_{\text{arb}}$.
The pseudocode of the extraction algorithm for $\tvis$ are listed in Algorithm \ref{A: T-VIS}.


\begin{algorithm}[tb]
    \caption{Extraction of Pruning Predicates for $\tvis$}
    \label{A: T-VIS}
    
    \KwIn{a set of abstract executions $\mathcal{A} = \{\pi_1, \pi_2 \ldots \pi_m\}$}
    \KwOut{a set of pruning predicates of $\tvis$}
    
    $S_{vis}$ $\leftarrow$ $\bigcup_{t=1}^{m} vis(\pi_t)$\\
    \ForEach{pair $(x,y) \in S_{vis}$}{
        $S_{arb}$ $\leftarrow$ $U$\\
        \ForEach{$\pi \in \mathcal{A}$ \textnormal{which} $(x,y) \in vis(\pi)$}{
            $S_{arb}$ $\leftarrow$ $S_{arb} \bigcap lin(\pi)$
        }
        
        \textit{review} $\leftarrow$ \textsf{TRUE}\\
        \ForEach{$\pi \in \mathcal{A}$ \textnormal{which} $(x,y) \notin vis(\pi)$}{
            \If{$S_{arb} \subseteq lin(\pi)$}{
                \textit{review} $\leftarrow$ \textsf{FALSE}\\
                \textbf{break}
            }
        }
        \If{\textit{review} \textnormal{is} \textnormal{\textsf{TRUE}}}{
            \textbf{yield} $\tvis(S_{{arb}}, x, y)$
        } 
    }
\end{algorithm}

Since the template $\tnotvis$ is just the dual of $\tvis$, the extraction of concrete predicates for $\tnotvis$ is similar to the extraction for $\tvis$. 
We just replace the visibility relation in the $\tvis$ extraction algorithm with the not-visibility relation. 
The pseudocode of the extraction algorithm for $\tnotvis$ are listed in Algorithm \ref{A: T-NOT-VIS}.

\begin{algorithm}[tb]
    \caption{Extraction of Pruning Predicates for $\tnotvis$}
    \label{A: T-NOT-VIS}
    
    \KwIn{a set of abstract executions $\mathcal{A} = \{\pi_1, \pi_2 \ldots \pi_m\}$}
    \KwOut{a set of pruning predicates of $\tnotvis$}
    
    $S_{notvis}$ $\leftarrow$ $\bigcup_{i=1}^{m} notvis(\pi_i)$\\
    \ForEach{pair $(x,y) \in S_{notvis}$}{
        $S_{arb}$ $\leftarrow$ $U$\\
        \ForEach{$\pi \in \mathcal{A}$ \textnormal{which} $(x,y) \in notvis(\pi)$}{
            $S_{arb}$ $\leftarrow$ $S_{arb} \bigcap lin(\pi)$
        }
        
        \textit{review} $\leftarrow$ \textsf{TRUE}\\
        \ForEach{$\pi \in \mathcal{A}$ \textnormal{which} $(x,y) \notin notvis(\pi)$}{
            \If{$S_{arb} \subseteq lin(\pi)$}{
                \textit{review} $\leftarrow$ \textsf{FALSE}\\
                \textbf{break}
            }
        }
        \If{\textit{review} \textnormal{is} \textnormal{\textsf{TRUE}}}{
            \textbf{yield} $\tnotvis(S_{arb}, x, y)$
        } 
    }
\end{algorithm}

\subsubsection{Correctness of Pruning}

We now prove that the pruning is safe, i.e., no certificate abstract execution will be falsely discarded by the pruning predicates. 
Here we present a sketch of the proof. 

Given a history $\mathcal{H}$ and a consistency model $\Phi$, if $\mathcal{H} \not\models \Phi$, it means that there exists no certificate abstract execution for $\Phi$. 
So the pruning must be safe. 
Therefore in the rest of the proof, we focus on the case in which history $\mathcal{H}$ has a certificate abstract execution satisfying the consistency model $\Phi$. 

The \vs algorithm prunes abstract executions during the construction of a candidate execution at runtime. 
So we need to prove that no candidate execution, which can be extended to a certificate abstract execution in the future, is pruned.
If an abstract execution violates a pruning predicate, all of its extensions violate the pruning predicate too. 
Hence, we only need to prove that there does not exist a certificate abstract execution which violates the pruning predicate. 

We assume for contradiction that there exists a certificate abstract execution $\pi$ of history $\mathcal{H}$ which violates a pruning predicate $P$.
The pruning predicate $P$ is extracted from some query cluster in history $\mathcal{H}$. 
We project the abstract execution $\pi$ on the query cluster and get the sub-execution $\pi'$. The sub-execution $\pi'$ is an abstract execution of the query cluster from which the predicate $P$ is extracted.
The arbitration and visibility relations among \textit{dictating updates} and \textit{dictated query} is reserved after projection. 
So $\pi'$ still violates the predicate $P$. 
Moreover, the \textit{dictated query} of the query cluster has the same return value after the projection. 

Since $\pi$ is a certificate abstract execution, each operation in $\pi$ can see all the operations it is required to see by the visibility constraint of $\Phi$. In other words, each operation in $\pi$ does not ignore any operation it is required to see. 
Compared to $\pi$, each operation in $\pi'$ only omits those operations discarded in the projection to a query cluster. 
It means that each operation in $\pi'$ does not ignore any operation it is required to see. 
%
Thus the projection $\pi'$ still satisfies the visibility constraint. This means that the projection $\pi'$ still satisfies the consistency model $\Phi$. 

We can also prove that the abstract execution $\pi'$ still satisfies the visibility constraint of the consistency model $\Phi$ in a formal way. 

As we construct the abstract execution by operation, the visibility relation of the first $k-1$ operations is fixed when we choose the visibility of the $k$-th operation $o_k$. 
We define a set $\mathcal{V}_k$ consisting of operations that is visible to the $k$-th operation. 
We extend the abstract execution $\ang{lin_{k-1}, vis_{k-1}}$ to $\ang{lin_{k}, vis_{k}}$ by adding 
$\mathcal{V}_k$ to $vis_{k-1}$ and adding $o_k$ to the end of $lin_{k}$. 

We define a function $\mathcal{G}$ mapping a tuple $(\ang{lin_{k-1}, vis_{k-1}}, \mathcal{V}_k, o_k)$ to a set $\mathcal{M}$. 
The set $\mathcal{M}$ is the visibility relation that describles what $o_k$ must see according to the visibility constraint of the consistency model. If $\mathcal{M} \subseteq \mathcal{V}_k$, $\mathcal{V}_k$ is a legal visibility extension for the consistency model $\Phi$. 
For example, we list some function $\mathcal{G}$ according to the visibility predicates in \ref{SubSec: Vis-Relax}, 
\begin{itemize}
    \item $\mathcal{G}_{weak}(\_, \_, o_k) = \emptyset$, which has no constraint for visibility. 
    \item $\mathcal{G}_{basic}(\_, \_, o_k) = \{(x, o_k)\ \vert\ \forall x: (x, o_k) \in so \}$, which means $o_k$ must see all the operations that happens before it according to the session order. 
    \item $\mathcal{G}_{monotonic}(\ang{lin_{k-1}, vis_{k-1}}, \_, o_k) = \{(x, o_k)\ \vert\ \forall x \exists y : (y, o_k) \in so \land (x, y) \in vis_{k-1}\}$, which means $o_k$ must see the operations that its predecessors see. For the sake of simplicity, each operation is visible to itself. 
    \item $\mathcal{G}_{causal}(\ang{lin_{k-1}, vis_{k-1}}, \mathcal{V}_k, o_k) =  \{(x, o_k)\ \vert\ \forall x :(x, o_k) \in (vis_{k-1} \bigcup \mathcal{V}_k)^{+}\} \bigcup \mathcal{G}_{basic}(\_, \_, o_k)$, which means $o_k$ must see operations visible to those $o_k$ sees, and the visibility is a transitive closure. 
    \item $\mathcal{G}_{complete}(\ang{lin_{k-1}, vis_{k-1}}, \_, o_k) = \{(x, o_k)\ \vert\ \forall x: (x, \_) or (\_, x) \in lin_{k-1} \}$
\end{itemize}

The set $\mathcal{R}$ contains relation pairs in the abstract execution $\pi$ but not in the projection $\pi$. 
$\mathcal{V}_{k'}$ is the visibility relation that describles operations visible to the $k$-th operation in $\pi$, i.e. the $k'$-th operation in $\pi'$ after projection.  
All relation pairs in $\mathcal{R}$ are removed from $\mathcal{V}_k$ after projection, so $\mathcal{V}_{k'} = \mathcal{V}_k \setminus \mathcal{R}$. 
Also operations in $\mathcal{R}$ are removed from $\mathcal{M}$ after projection. Considering the transitivity, some operations cannot see another operation via removed operations in $\mathcal{R}$ any more, 
\begin{equation*}
    \begin{split}
        &\mathcal{M}' = \mathcal{G}(\ang{lin_{k'-1}, vis_{k'-1}}, \mathcal{V}_{k'}, o_{k'}) \\
        &\subseteq \mathcal{G}(\ang{lin_{k-1}, vis_{k-1}}, \mathcal{V}_k, o_k) \setminus \mathcal{R} = \mathcal{M} \setminus \mathcal{R}.
    \end{split}
\end{equation*} 

As the original abstract execution $\pi$ satisfies the visibility constraint of the consistency model $\Phi$, we can conclude that
$\mathcal{M} \subseteq \mathcal{V}_k$ and further deduce that $\mathcal{M} \setminus \mathcal{R} \subseteq \mathcal{V}_k \setminus \mathcal{R}$. 
Combining the transitivity of $\subseteq$, we can conclude that 
$\mathcal{M}' \subseteq \mathcal{V}_{k'}$. 
Therefore, the projection $\pi'$ still satisfies the visibility constraint of the consistency model $\Phi$. 

Because the pruning predicate $P$ is extracted from all valid abstract executions of the query cluster, each valid abstract execution of the query cluster must satisfy predicate $P$ according to the extraction algorithm. 
This leads to contradiction, since we have $\pi'$ must satisfy $P$ and violate $P$ at the same time.
Therefore, the pruning in \vs is safe, in the sense that the candidate abstract executions discarded can never be extended to a valid certificate one.

\subsection{Parallelization of \vs}

\begin{figure}[tb]
    \includegraphics[width=0.9\linewidth]{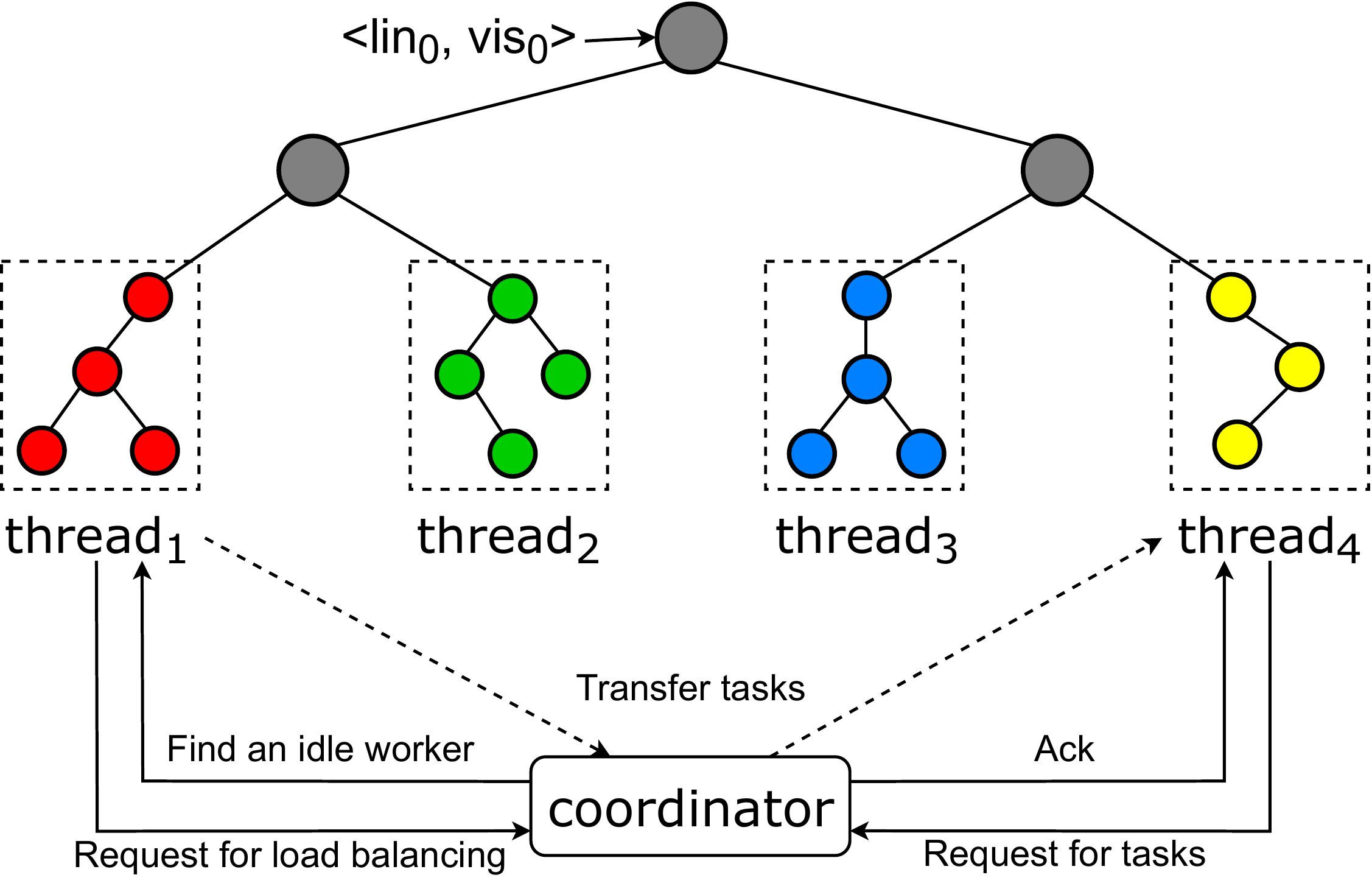}
    \centering
    \caption{Parallel \vs}
    \label{F: Parallel}
\end{figure}

All legal search states of a history are organized as a tree, as shown in Fig.~\ref{F: Parallel}. 
Each search state is a node in the tree, and the root node is the empty abstract execution $\ang{lin_0, vis_0}$.
For a node $\ang{lin, vis}$ in the tree, search states which are possible extensions of $\ang{lin, vis}$ in the next phase are the children of $\ang{lin, vis}$. 
There may exist many candidate extensions in one phase, and different choices of extensions result in the search of different subtrees. 

Consistency checking with \vs is principally a depth-first search, aiming to find a search state which constructs a certificate abstract execution.
Therefore, there exists natural parallelization for the search. 
Subtrees do not share any nodes and are independent of each other. 
We can search the subtrees in parallel. 
If we find a certificate execution in any subtree, we terminate the search. 
If no certificate executions could be found in any subtree, the history does not satisfy the consistency model. 

In preparation for the parallel search, the central coordinator runs the sequential \vs algorithm and generates a batch of search states. 
These search states are evenly allocated to different threads. Then each thread runs an instance of the sequential \vs algorithm.
Note that here we use the term ``thread" in a broad sense. The thread is more like an abstract worker. It can also be, for example, a virtual or physical machine.
As shown in the example in Fig.~\ref{F: Parallel}, we generate four search states (the root node of each subtree in the dashed box). 
Each of the four states is allocated to a worker thread. The four threads extend the search states in parallel and obtain four different subtrees (the colored subtrees in four dashed boxes).

Since the search cost for different subtrees may vary quite a lot, we further design a load balancing scheme for the coordinator.
Specifically, when a thread finishes its search task, it asks the coordinator for more tasks.
Meanwhile, the fully loaded threads periodically contact the coordinator to see whether there are idle threads which they can share their workloads with.
In this way, the working thread can allocate part of its own tasks to other threads via the coordinator.
Since the thread maintains its tasks as search states in the deque (see detailed design in Section \ref{SubSubSec: Revised-VS}), it simply transfers part of its workloads from the tail of its deque to other threads via the coordinator.

We let the working threads periodically contact the coordinator, instead of letting the coordinator interrupt the working threads.
This is mainly because interrupting a running thread usually imposes more overhead.
Based on the statistics from running \vs without load balancing, we let the working thread contact the coordinator every 1000 search states in our experiments in the following Section \ref{Sec: Exp}.

\section{Experimental Evaluation} \label{Sec: Exp}

In this section, we apply the \vs framework to consistency measurement in two open-source replicated data type stores.
The experiments are aimed at two research questions:

\begin{itemize}
    \item RQ1. Can the \vs framework enable precise measurement of consistency semantics for eventually consistent replicated data types?
    \item RQ2. Is the consistency measurement based on \vs efficient?
\end{itemize}

%

\noindent We first present our experiment setup and design, and then discuss the evaluation results.
The implementation of the \vs framework, as well as implementations, configurations and experiment scripts over the two replicated data type stores are available in our online GitHub repository \cite{ViSearch-GitHub}.

\subsection{Experiment Setup}

We employ \vs to conduct consistency measurement for Riak \cite{Riak} and CRDT-Redis \cite{CRDT-Redis}. 
Riak is a distributed NoSQL key-value database, which provides Riak-specific data types based on CRDTs. 
CRDT-Redis revises Redis, a popular in-memory data structure store with built-in replication, into a multi-master architecture and supports CRDT-based conflict resolution for concurrent updates on multiple master nodes.

We select three widely-used data container types, \textit{set}, \textit{map}, and \textit{priority queue}.
We further configure each container type we select as follows.
Riak exposes a parameter $r$ to the clients, which denotes how many replicas a query request must contact.
The two data types with different parameters $r$ $(r=1 \text{ or } 2$) are denoted \textit{Set-r1}, \textit{Set-r2}, \textit{Map-r1} and \textit{Map-r2} in our experiment. 
Each data type in CRDT-Redis is designed with different conflict-resolution strategies: add-win (\textit{A} in short), remove-win (\textit{R} in short), and the \rwf framework (\textit{F} in short) \cite{Zhang21}. 
Data types with different conflict-resolve strategies are denoted \textit{Set-A}, \textit{Set-R}, \textit{Set-F}, \textit{PQ-A}, \textit{PQ-R}, and \textit{PQ-F}. 

We run both data type stores in the Alibaba Cloud Elastic Compute Service (ECS) \cite{Alibaba-Cloud}. Each ECS server has two Intel Xeon Platinum CPU (2.50GHz), 4GB RAM, 40GB ESSD and 1MB/s network bandwidth. We build a Riak cluster with 5 servers, and replicate data types on 3 servers in the cluster. The CRDT-Redis cluster consists of 3 servers, and data types are replicated on all 3 servers. Another ECS server is dedicated to running the clients. 
The checking of data type access traces using \vs is conducted on one workstation with an Intel i9-9900X CPU (3.50GHz), with 10 cores and 20 threads, and 32GB RAM, running Ubuntu Desktop 16.04.6 LTS.

\begin{figure}[tb]
    \centering
    \includegraphics[width=0.9\linewidth]{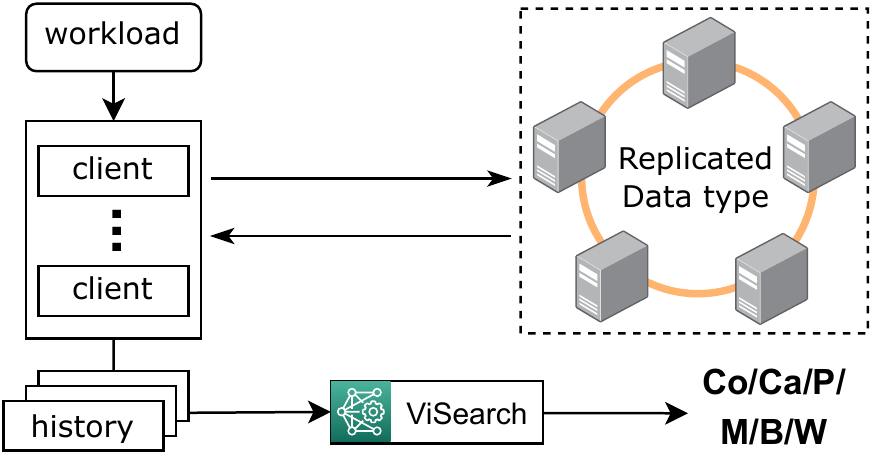}
    \caption{\vs workflow.}
    \label{F: Exp}
\end{figure}

The basic workflow of consistency measurement using \vs is shown in Fig.~\ref{F: Exp}.
We first generate the workloads for each data type. 
The clients record both the query and update invocation info, 
and the return values 
for all queries issued.
All the info of data type access collected by the clients in one experiment forms one trace or history of data type access. 
Multiple traces of data type access are imported into the \vs module and \vs tells us the consistency levels of the traces measured. 
Detailed definitions of the consistency levels can be found in Section \ref{SubSec: Vis-Relax}.


\subsection{Experiment Design} \label{SubSec: Exp-Design} 

The optimized trace checking using \vs (as detailed in Section \ref{Sec: VS}) still requires exponential time in the worst case. Thus we cannot directly apply \vs over CRDT access traces of arbitrary length. As inspired by the measurement in \cite{Emmi19}, we explore locality in consistency measurement over CRDT traces to circumvent the NP-hardness obstacle.

As for a CRDT, any replica can be modified without coordination with other replicas. Thus the replicas  may temporarily diverge. 
We call the period when the replicas diverge the \textit{divergence interval}. 
Our key observation for consistency measurement is that, the consistency level of given trace is decided by the divergence intervals in the trace. Thus we only need to measure the divergence intervals in the trace, instead of measuring the whole trace of arbitrary length. The consistency level of the whole trace is decided by the divergence interval resulting in the weakest consistency level.
We can estimate the divergence interval by monitoring how large the number of updates in transmission can be. 
%
%
%
We find that the number of operations in transmission is bounded by 15 in our experiments. 
Therefore we set the size of the workload, i.e., the number of operations to be sent to the data type, to 15. 
We let the clients launch the operations as quickly as possible to produce the maximal divergence.
Note that the specific value of the divergence interval is not essential in our experiment. 
In different experiment scenarios, we can estimate the divergence interval in the same way. 

In our workloads, about 60\% operations are \textit{updates}, and about 40\% \textit{queries}. 
We randomly choose the parameters from a small interval of 0$\sim$5, which is no more than 1/3 of the workload size. 
This ensures that there exists sufficient collisions among operations.
As discussed above, each workload contains about 15$\sim$17 operations and 3$\sim$5 client threads share the operations in a workload. 

Note that the consistency measurement based on \vs is an approximation of the actual consistency semantics provided by the data type, since \vs cannot traverse the infinite space of possible traces.
To ensure the accuracy of our consistency measurement, we decide whether we have checked a sufficient number of traces in the Las Vegas manner \cite{Motwani95}.
We keep running \vs over the data type access traces round by round.
In each round, we check 100,000 traces. 
According to our experience in CRDT implementation and testing, 100,000 traces are large enough to cover all corner cases in the CRDT design.
%
The weakest consistency level measured over all the 100,000 traces is the measurement result in this round.
We stop the measurement when the measure results remain unchanged in 3 consecutive rounds.

\subsection{Evaluation Results} \label{SubSec: Eval-Results}

In light of our two research questions, we first discuss what the application developers and CRDT designers can obtain from our consistency measurement. 
Then we study the checking cost. 
We also study the effects of the two optimization techniques we develop. 

\subsubsection{Consistency Measurement Results}

We measure the consistency levels of 10 replicated data types from two data type stores. 
Due to different data type configurations and conflict resolution strategies, these replicated data types provide different consistency levels. 
We expect \vs can differentiate the consistency semantics of the 10 data types we study.
The measurement results are listed in Table. \ref{T: Result}. 
Each row of the table corresponds to one data type.
For each of the 6 consistency levels, we list the number of violations found by \textsc{ViSearch}.
The strongest consistency level with 0 violation (colored gray in the table) is the consistency level of the data type.
The `/' means that the checking is unnecessary and the violation must be zero, since there is no violation for stronger consistency levels.

We demonstrate the usefulness of the consistency measurement results from the perspective of the ``consistency-latency tradeoff", which is a refinement of the widely-used CAP theorem \cite{Brewer12, Gilbert12} and the PACELC tradeoff \cite{Abadi12} in CRDT scenarios.
Different CRDT designs and implementations principally have the same data update delay, since the replica is updated locally and the updates are propagated asynchronously. The query delay is also principally the same if the query only contacts one replica. In some data stores like Riak, the client can choose to contact multiple replicas, sacrificing certain query delay and improving the consistency level of the query.
Thus, from the perspective of the consistency-latency tradeoff, it is the consistency semantics of the data type which captures the primary differences in CRDT designs and implementations.
%

The consistency measurement of CRDT-Redis data types demonstrates the different consistency semantics resulting from different conflict resolution strategies.
The priority queue \textit{PQ-R} with the remove-win strategy merely provides the ``Weak" consistency level, while the priority queue \textit{PQ-A} with the add-win strategy and \textit{PQ-F} with the \rwf framework both provide the stronger consistency level ``peer".
Moreover, the number of violations of the remove-win strategy is much larger than that of the other two strategies on the causal consistency level.

Considering the high level design rationale, we may reasonably think that the remove-win strategy and the add-win strategy are dual of each other and are principally the same. However, the measurement by \vs shows important differences in the consistency semantics. 
The \rwf framework is an improvement on the traditional remove-win strategy. The remove-win strategy puts the remove operation behind the non-remove operation. In \textsc{RWF}, the remove-operation just removes the element from the container, thus avoiding the conflict. This \rwf strategy, seen as an improvement of the traditional remove-win strategy, strengthens the consistency semantics to the same level with the semantics of the add-win strategy.
The consistency measurement results can help the CRDT designers to further understand the semantics of different resolution strategies. The application developers can also use the measurement results to choose appropriate data types they actually need.

The consistency measurement results of the set data type have important differences with those of the priority queue data type.
As for the add-win and \rwf resolution strategies, the consistency level of \textit{Set-A} and \textit{Set-F} is still ``Peer", which is the same with the level of their counterparts \textit{PQ-A} and \textit{PQ-F}.
The main difference appears when considering the remove-win strategy. As for the set type, \textit{Set-R} also provides ``Peer", while \textit{PQ-R} only provides ``Weak".
The difference mainly results from that the priority queue has much more complex semantics than the set. When coping with complex semantics of the data type, the remove-win strategy will decrease the level of data consistency provided to the upper-layer applications. 
These measurement results show that the semantics of the data type should also be considered when choosing conflict resolution strategies. 

As for the Riak case, the configuration of parameter $r$ plays the most important role in the consistency measurement. The effect of $r$ can be quantified by the consistency measurement based on \textsc{ViSearch}.
We find that both data types merely provide the ``Weak" consistency level when $r=1$.
When $r=2$, a query request must contact a majority of replicas before returning. 
According to quorum-based protocol, data types with $r=2$ cannot strictly guarantee strong consistency. 
However, \textit{Set-r2} and \textit{Map-r2} actually provide strong consistency (the ``Complete" consistency level) in the experiment.
The measurement results show that when the network condition is relatively reliable and the communication delay is low, e.g., in an intra-datacenter environment, the ``$r=2$" configuration actually provides strong consistency, at least with high probability.
This result is quite useful for applications which want strong consistency, but can tolerate data inconsistency in rare cases \cite{Wei17, Ouyang21}. 

\begin{table}[tb]
    \caption{Consistency measurement results.}
    \label{T: Result}
    \resizebox{\linewidth}{!}{  
        \begin{tabular}{|c|cccccc|l|l|}
            \hline
            & \multicolumn{6}{c|}{Violations}       &       &       \\ \cline{2-7}
            \multirow{-2}{*}{\begin{tabular}[c]{@{}c@{}}Data \\ Type\end{tabular}} & \multicolumn{1}{c|}{Co}            & \multicolumn{1}{c|}{Ca}       & \multicolumn{1}{c|}{P}        & \multicolumn{1}{c|}{M}        & \multicolumn{1}{c|}{B}        & W         & \multirow{-2}{*}{\begin{tabular}[c]{@{}l@{}}\# of \\ Histories\end{tabular}} & \multirow{-2}{*}{\begin{tabular}[c]{@{}l@{}}Total \\ Time\end{tabular}} \\ \hline
            \textit{Set-r1}     & \multicolumn{1}{c|}{366}                      & \multicolumn{1}{c|}{339}                      & \multicolumn{1}{c|}{334}                              & \multicolumn{1}{c|}{295}       & \multicolumn{1}{c|}{188}       & \cellcolor[gray]{.8}0     & 300k          & 8h 3min      \\ \hline
            \textit{Set-r2}     & \multicolumn{1}{c|}{\cellcolor[gray]{.8}0}    & \multicolumn{1}{c|}{/}                        & \multicolumn{1}{c|}{/}                                & \multicolumn{1}{c|}{/}        & \multicolumn{1}{c|}{/}        & /                         & 300k          & 0h 11min      \\ \hline
            \textit{Map-r1}     & \multicolumn{1}{c|}{788}                      & \multicolumn{1}{c|}{734}                      & \multicolumn{1}{c|}{732}                                & \multicolumn{1}{c|}{712}    & \multicolumn{1}{c|}{542}      & \cellcolor[gray]{.8}0     & 300k          & 24h 26min       \\ \hline         
            \textit{Map-r2}     & \multicolumn{1}{c|}{\cellcolor[gray]{.8}0}    & \multicolumn{1}{c|}{/}                        & \multicolumn{1}{c|}{/}                                & \multicolumn{1}{c|}{/}        & \multicolumn{1}{c|}{/}        & /                         & 300k          & 0h 11min      \\ \hline
            \textit{PQ-A}       & \multicolumn{1}{c|}{478}                      & \multicolumn{1}{c|}{3}                        & \multicolumn{1}{c|}{\cellcolor[gray]{.8}0}            & \multicolumn{1}{c|}{/}        & \multicolumn{1}{c|}{/}        & /                        & 300k          & 4h 49min       \\ \hline                
            \textit{PQ-R}       & \multicolumn{1}{c|}{704}                      & \multicolumn{1}{c|}{131}                        & \multicolumn{1}{c|}{130}            & \multicolumn{1}{c|}{119}        & \multicolumn{1}{c|}{119}        & \cellcolor[gray]{.8}0                        & 300k          & 14h 17min       \\ \hline                
            \textit{PQ-F}       & \multicolumn{1}{c|}{1153}                      & \multicolumn{1}{c|}{3}                        & \multicolumn{1}{c|}{\cellcolor[gray]{.8}0}            & \multicolumn{1}{c|}{/}        & \multicolumn{1}{c|}{/}        & /                        & 300k          & 0h 18min       \\ \hline                
            \textit{Set-A}      & \multicolumn{1}{c|}{45}                        & \multicolumn{1}{c|}{1}    & \multicolumn{1}{c|}{\cellcolor[gray]{.8}0}                                & \multicolumn{1}{c|}{/}        & \multicolumn{1}{c|}{/}        & /                         & 500k          & 0h 19min         \\ \hline                
            \textit{Set-R}      & \multicolumn{1}{c|}{15}                        & \multicolumn{1}{c|}{4}    & \multicolumn{1}{c|}{\cellcolor[gray]{.8}0}                                & \multicolumn{1}{c|}{/}        & \multicolumn{1}{c|}{/}        & /                         & 400k          & 8h 9min         \\ \hline                
            \textit{Set-F}      & \multicolumn{1}{c|}{5}                        & \multicolumn{1}{c|}{1}    & \multicolumn{1}{c|}{\cellcolor[gray]{.8}0}                                & \multicolumn{1}{c|}{/}        & \multicolumn{1}{c|}{/}        & /                         & 500k          & 0h 19min         \\ \hline                
    \end{tabular}}
\end{table}

\subsubsection{Consistency Measurement Cost}

To obtain the consistency measurement results in Table \ref{T: Result}, the checking time of \vs varies from 11 minutes to 24 hours. 
The consistency measurement time varies significantly for different data types. In our experiment 4 out of 10 types, \textit{Set-r1}, \textit{Map-r1}, \textit{PQ-R}, and \textit{Set-R}, require significantly more time than other data types. Thus for most data types, the number of traces checked can be increased to a much larger number (e.g., 1 million) in reasonable time. 
The checking time results demonstrate that consistency measurement using \vs can be accomplished under reasonable testing budget.
Also note that, the measurement process is not on the critical path of CRDT design, implementation, testing and maintenance. The consistency measurement can ``borrow" the traces from the testing process and be finished in parallel with the testing of CRDT implementations.

\begin{figure}[tb]
    \centering
    \subfigure[PQ-F] {
        \includegraphics[width=0.46\hsize]{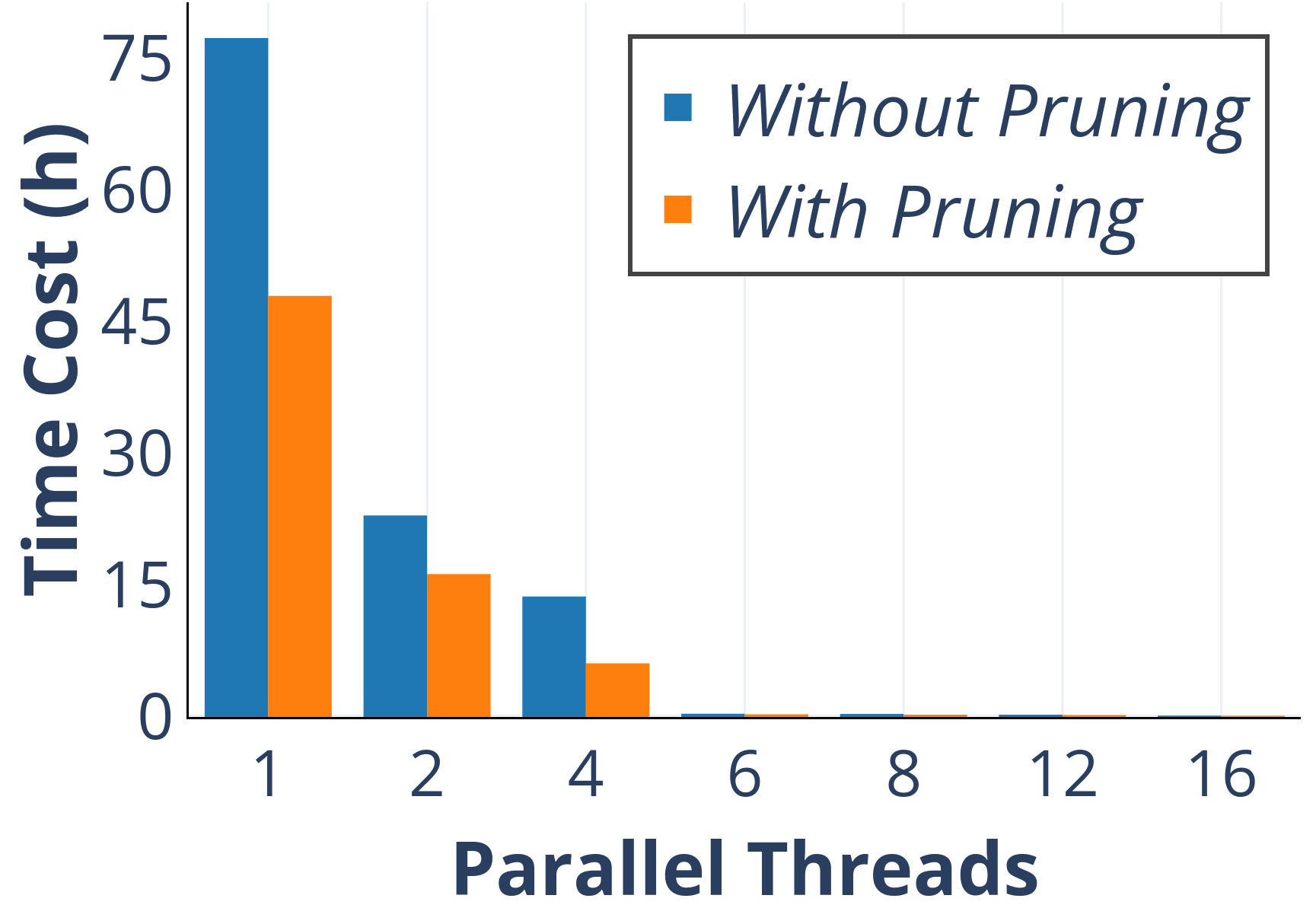}
    }
    \subfigure[Set-r1] {
        \includegraphics[width=0.46\hsize]{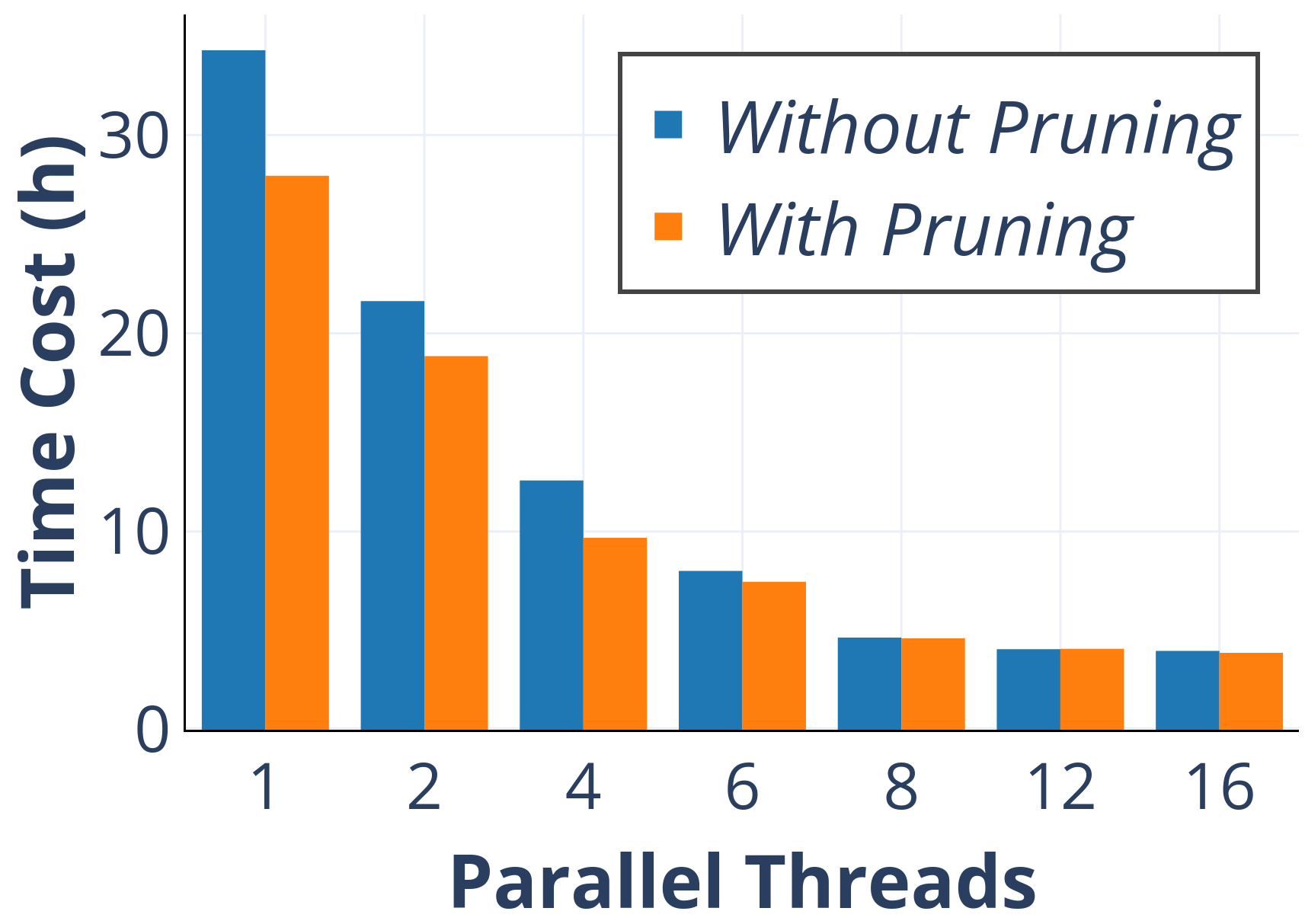}
    }
    \caption{Evaluation of Time Cost.}
    \label{parallel}
\end{figure}

Besides the total checking time using \textsc{ViSearch}, we further study the effects of the optimizations we propose, namely the pruning of search space and the parallelization of checking.
We use \vs without pruning and parallelization as our baseline.
As for the parallelization of consistency measurement, we study two data types \textit{Set-r1} and \textit{PQ-F}.
We measure 100,000 traces and change the number of threads of checking from 1 to 16. 
As shown in Fig.~\ref{parallel}, the checking time decreases as we increase the number of threads, using pruning or not.
The decrease in time is quite significant when we increase the thread number from 1 to 6 in the cases of \textit{PQ-F} and \textit{Set-r1}. 
As we add more threads, the checking time decreases approximately linearly. This means that the time decreases slowly than the ideal case where the time decreases to $\frac{1}{k}$ $(k$ is the number of threads).
This is mainly because the cost for thread scheduling increases when the thread number increases, which impacts the efficiency of parallelization. 
Also, when the thread number increases, the load balancing among threads becomes harder and more overloading/starvation of the threads will impact the efficiency of parallelization.
This evaluation results can tell us what the sweet points are, in order to choose the appropriate number of threads. 
It also motivates us to design better schemes to divide workloads among threads more efficiently.

As for the pruning, we use the \textit{pruning ratio} as the performance metric. 
As for the checking of one data type access trace, we count the number of explored search states. 
We calculate the \textit{pruning ratio}, i.e. the ratio of the number of explored search states checked with the pruning to the number of states explored without the pruning.
For data types studied in the experiment, we pick one representative from multiple configurations of this data type. Specifically, we calculate the pruning ratio of four data types: \textit{Set-r1}, \textit{Map-r1}, \textit{Set-F}, and \textit{PQ-F}. 

\begin{figure}[tb]
    \centering
    \includegraphics[width=\linewidth]{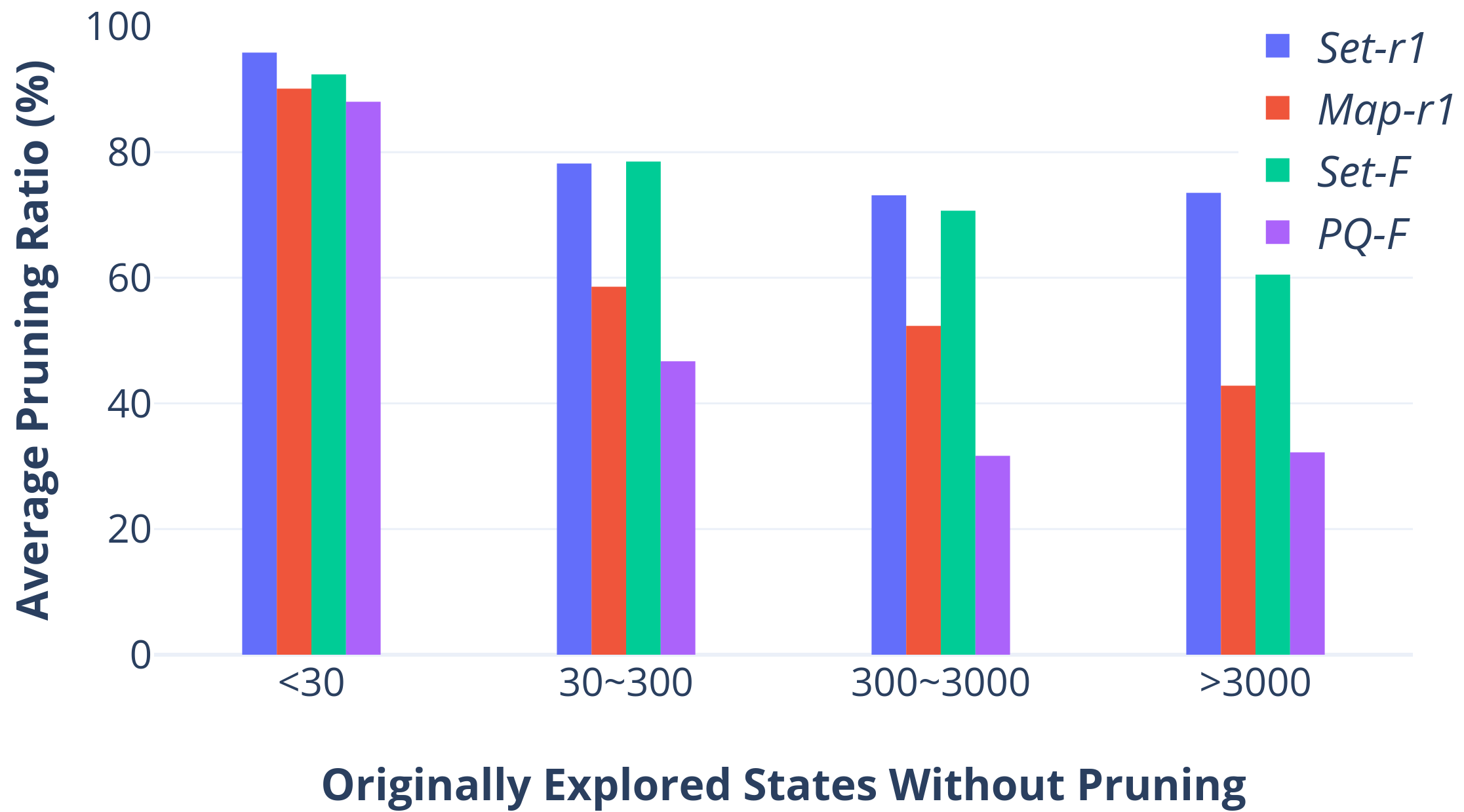}
    \caption{Evaluation of State Filter}
    \label{F: Pruning}
\end{figure}

As for the pruning ratio, we find that the effects of pruning significantly vary, as the number of search states (without pruning) increases. Our intuition is that the effects of pruning are more significant when there are more states to be explored by \textsc{ViSearch}, and the evaluation results prove this intuition correct.
Specifically, we divide the number of explored states without the pruning into four grades: \textit{small} ((0,30]), \textit{moderate} ([31, 300]), \textit{large} ([301, 3000]) and \textit{huge} ([3001, $+\infty$]).
As shown in Fig.~\ref{F: Pruning}, except for the `small' grade, the pruning can reduce the number of search states approximately by half. Note that, for `small' traces, the checking time is quite limited, with or without the pruning. This shows that, for the total checking time using the \vs framework, the pruning can significantly reduce the checking time (approximately by half).

One exception is \textit{Set-r1}. The pruning ratio of \textit{Set-r1} is more than 70\%, even for huge traces. This is mainly because \textit{Set-r1} provides weak visibility (see Table \ref{T: Result}), and the weak visibility checking takes most of the time. However, the weak visibility consistency predicate imposes few constraints on the visibility relation. We can find few pruning predicates conforming to the predicate template over all possible abstract executions. 
Therefore, the effect of pruning for \textit{Set-r1} is limited. 
We can further find that the pruning does not work well for the set type. Because the semantics of set only cares about the existence of elements and imposes fewer constraints on operations.
So we find fewer pruning predicates for the set type. 

Besides the pruning ratio, we also directly compare the checking time with and without the pruning.
Note that the checking cost with pruning (the orange bars in Fig.~\ref{parallel}) includes the time for the extraction of pruning predicates.
In Fig.~\ref{F: Pruning}, \textit{PQ-F} has a better pruning ratio than \textit{Set-r1}, and this is reflected in the time cost. In Fig.~\ref{parallel}, the pruning also has a better effect in \textit{PQ-F} than in \textit{Set-r1}. The pruning reduces the checking time by half in most cases for \textit{PQ-F}, while the pruning reduces the checking time by less than 30\% for \textit{Set-r1}.

We also find in Fig.~\ref{parallel} that the effect of pruning becomes less significant as the number of threads increases.
Note that there is a sharp decrease in checking time as the number of threads goes beyond 4 in the \textit{PQ-F} case. 
It costs 0.36(0.30), 0.35(0.26), 0.26(0.21), 0.16(0.14) hours without(with) pruning when the number of parallel threads is 6, 8, 12, and 16.
The parallelization negatively affects the effect of the pruning, mainly because if any worker finds a certificate abstract execution, the searching process can be directly terminated.
Parallel checking greatly increases the odds of this type of early termination (in our searching, we can find some certificate abstract execution in most cases).
In light of the increased odds of early termination, the effects of pruning is significantly restricted as the number of worker threads increases.

\section{Related Work} \label{Sec: RW}


The fuzzy nature of eventual consistency and its wide adoption motivate the quantification of severity of consistency violations \cite{Golab11, Golab14, Golab18}. Existing quantification techniques can be viewed from two perspectives. From the micro-perspective, existing consistency measurement schemes are mainly based on certain metrics delineating the staleness of data, e.g., how long the data has been stale and how many new versions of data have been generated.
Time-based quantification techniques, e.g. $\Delta$-atomicity \cite{Golab11} and $\Gamma$-atomicity \cite{Golab14}, are adopted for applications sensitive to realtime consistency requirements. Time-based techniques measure data consistency by how many time units the value returned has been stale. 
For applications sensitive to the frequency of data updates, version-based quantification techniques , e.g. k-atomicity \cite{Golab18}, are proposed. Version-based techniques measure data consistency by how many times the data has been updated when the returned value is obtained.
From the macro-perspective, Yu and Vahdat propose TACT (Tunable Availability and Consistency Tradeoffs), a middleware layer that supports application-dependent consistency
semantics expressed as a vector of metrics defined over a logical consistency unit or conit \cite{Yu02-TOCS}. Inconsistencies in the observed value of a conit are bounded using three metrics: numerical error, order error, and staleness.

The consistency measurement schemes above are from the end user's perspective. In order to better facilitate the development of upper-layer applications, we need measurement schemes which focus on how data items are updated, in a similar way with how linearizability is specified. To this end, in this work we borrow the \visar framework which is an axiomatic consistency specification framework of concurrent objects. 
The \visar framework extends the linearizability-based specification methodology with a dynamic visibility relation among operations, in addition to the standard dynamic happens-before and linearization relations. 
The \visar framework is more suitable for developers to reason about the correctness of the implementation of a replicated data type. It is also suitable for upper-layer application developers to reason about the correctness of applications.


The checking of data type access traces is NP-Complete in general \cite{Gibbons97}.
However, this intractability result can be circumvented by reasonable assumptions in specific scenarios. For example, the write values are often assumed to be unique 
\cite{Gibbons97}. This assumption is reasonable and easy to satisfy, since we can append certain meta-data to the values written to make them unique. 
In replicated data type store scenarios, the number of concurrent writes is often assumed to be bounded by some constant \cite{Golab18}. 
Practitioners can also limit the pattern of the input and derive an input specific efficient checking algorithm \cite{Bouajjani17}. 

The checking of data type access traces using \vs is also intractable in the worst case. As inspired by the existing techniques to circumvent the NP-hardness obstacle, we refactor the brute-force checking algorithm to a generic algorithm skeleton. This refactor enables practical pruning of the exponential search space. We also employ the locality of the checking of CRDT access traces to make the consistency measurement practical in realistic replicated data type store scenarios.


\section{Conclusion and Future Work} \label{Sec: Concl}

In this work, we develop the \vs framework for consistency measurement of eventually consistent replicated data type stores. 
In our framework, different consistency levels are specified via the relaxation of the visibility relations among operations.
The measurement of data consistency is based on the \vs algorithm skeleton, which enables efficient pruning of the search space and effective parallelization.
We employ the \vs framework for consistency measurement in two replicated data type stores Riak and CRDT-Redis. The experimental evaluation shows the usefulness and cost-effectiveness of consistency measurement based on the \vs framework in realistic scenarios.

Currently, the \vs framework is being integrated into the Jepsen framework \cite{Jepsen-GitHub} for black-box testing of distributed data stores (see details in the online repository of \vs \cite{ViSearch-GitHub}).
We will conduct more case studies to further demonstrate the usefulness of \vs on the basis of Jepsen testing.
In our future work, we will also investigate how to propose more predicate templates and extract efficient pruning predicates to further reduce the consistency measurement cost. 
Moreover, efficient heuristics can also be employed to reduce the search cost.


\balance


\bibliographystyle{ACM-Reference-Format}
\bibliography{visearch}


\begin{thebibliography}{32}


\ifx \showCODEN    \undefined \def \showCODEN     #1{\unskip}     \fi
\ifx \showDOI      \undefined \def \showDOI       #1{#1}\fi
\ifx \showISBNx    \undefined \def \showISBNx     #1{\unskip}     \fi
\ifx \showISBNxiii \undefined \def \showISBNxiii  #1{\unskip}     \fi
\ifx \showISSN     \undefined \def \showISSN      #1{\unskip}     \fi
\ifx \showLCCN     \undefined \def \showLCCN      #1{\unskip}     \fi
\ifx \shownote     \undefined \def \shownote      #1{#1}          \fi
\ifx \showarticletitle \undefined \def \showarticletitle #1{#1}   \fi
\ifx \showURL      \undefined \def \showURL       {\relax}        \fi
\providecommand\bibfield[2]{#2}
\providecommand\bibinfo[2]{#2}
\providecommand\natexlab[1]{#1}
\providecommand\showeprint[2][]{arXiv:#2}

\bibitem[\protect\citeauthoryear{??}{Ria}{2021}]%
        {Riak}
 \bibinfo{year}{2021}\natexlab{}.
\newblock
\newblock
\urldef\tempurl%
\url{https://github.com/basho/riak}
\showURL{%
Retrieved Dec 20, 2021 from \tempurl}


\bibitem[\protect\citeauthoryear{??}{Red}{2021}]%
        {Redis-Enterprise}
 \bibinfo{year}{2021}\natexlab{}.
\newblock
\newblock
\urldef\tempurl%
\url{https://redis.com/redis-enterprise/}
\showURL{%
Retrieved Dec 27, 2021 from \tempurl}


\bibitem[\protect\citeauthoryear{??}{Cos}{2021}]%
        {CosmosDB}
 \bibinfo{year}{2021}\natexlab{}.
\newblock
\newblock
\urldef\tempurl%
\url{https://azure.microsoft.com/en-us/services/cosmos-db/}
\showURL{%
Retrieved Dec 27, 2021 from \tempurl}


\bibitem[\protect\citeauthoryear{??}{CRD}{2021}]%
        {CRDT-Redis}
 \bibinfo{year}{2021}\natexlab{}.
\newblock
\newblock
\urldef\tempurl%
\url{https://github.com/elem-azar-unis/CRDT-Redis}
\showURL{%
Retrieved Dec 20, 2021 from \tempurl}


\bibitem[\protect\citeauthoryear{??}{Ali}{2021}]%
        {Alibaba-Cloud}
 \bibinfo{year}{2021}\natexlab{}.
\newblock
\newblock
\urldef\tempurl%
\url{https://www.alibabacloud.com}
\showURL{%
Retrieved Dec 20, 2021 from \tempurl}


\bibitem[\protect\citeauthoryear{??}{ViS}{2022}]%
        {ViSearch-GitHub}
 \bibinfo{year}{2022}\natexlab{}.
\newblock
\newblock
\urldef\tempurl%
\url{https://github.com/ViSearch/ViSearch}
\showURL{%
Retrieved Jun 18, 2022 from \tempurl}


\bibitem[\protect\citeauthoryear{??}{Jep}{2022}]%
        {Jepsen-GitHub}
 \bibinfo{year}{2022}\natexlab{}.
\newblock
\newblock
\urldef\tempurl%
\url{https://github.com/jepsen-io/jepsen}
\showURL{%
Retrieved Jul 5, 2022 from \tempurl}


\bibitem[\protect\citeauthoryear{Abadi}{Abadi}{2012}]%
        {Abadi12}
\bibfield{author}{\bibinfo{person}{Daniel~J. Abadi}.}
  \bibinfo{year}{2012}\natexlab{}.
\newblock \showarticletitle{Consistency Tradeoffs in Modern Distributed
  Database System Design: CAP is Only Part of the Story}.
\newblock \bibinfo{journal}{\emph{Computer}} \bibinfo{volume}{45},
  \bibinfo{number}{2} (\bibinfo{date}{Feb} \bibinfo{year}{2012}),
  \bibinfo{pages}{37--42}.
\newblock
\urldef\tempurl%
\url{https://doi.org/10.1109/MC.2012.33}
\showDOI{\tempurl}


\bibitem[\protect\citeauthoryear{Bailis, Venkataraman, Franklin, Hellerstein,
  and Stoica}{Bailis et~al\mbox{.}}{2014}]%
        {Bailis14-vldbj}
\bibfield{author}{\bibinfo{person}{Peter Bailis}, \bibinfo{person}{Shivaram
  Venkataraman}, \bibinfo{person}{Michael~J. Franklin},
  \bibinfo{person}{Joseph~M. Hellerstein}, {and} \bibinfo{person}{Ion Stoica}.}
  \bibinfo{year}{2014}\natexlab{}.
\newblock \showarticletitle{Quantifying eventual consistency with {PBS}}.
\newblock \bibinfo{journal}{\emph{{VLDB} J.}} \bibinfo{volume}{23},
  \bibinfo{number}{2} (\bibinfo{year}{2014}), \bibinfo{pages}{279--302}.
\newblock


\bibitem[\protect\citeauthoryear{Bermbach}{Bermbach}{2014}]%
        {Bermbach14}
\bibfield{author}{\bibinfo{person}{David Bermbach}.}
  \bibinfo{year}{2014}\natexlab{}.
\newblock \emph{\bibinfo{title}{Benchmarking Eventually Consistent Distributed
  Storage Systems}}.
\newblock \bibinfo{thesistype}{Ph.D. Dissertation}. \bibinfo{school}{Karlsruhe
  Institute of Technology}.
\newblock


\bibitem[\protect\citeauthoryear{Bouajjani, Enea, Guerraoui, and
  Hamza}{Bouajjani et~al\mbox{.}}{2017}]%
        {Bouajjani17}
\bibfield{author}{\bibinfo{person}{Ahmed Bouajjani},
  \bibinfo{person}{Constantin Enea}, \bibinfo{person}{Rachid Guerraoui}, {and}
  \bibinfo{person}{Jad Hamza}.} \bibinfo{year}{2017}\natexlab{}.
\newblock \showarticletitle{On Verifying Causal Consistency}. In
  \bibinfo{booktitle}{\emph{Proceedings of the 44th ACM SIGPLAN Symposium on
  Principles of Programming Languages}} (Paris, France)
  \emph{(\bibinfo{series}{POPL 2017})}. \bibinfo{publisher}{Association for
  Computing Machinery}, \bibinfo{address}{New York, NY, USA},
  \bibinfo{pages}{626–638}.
\newblock
\showISBNx{9781450346603}
\urldef\tempurl%
\url{https://doi.org/10.1145/3009837.3009888}
\showDOI{\tempurl}


\bibitem[\protect\citeauthoryear{{Brewer}}{{Brewer}}{2012}]%
        {Brewer12}
\bibfield{author}{\bibinfo{person}{E. {Brewer}}.}
  \bibinfo{year}{2012}\natexlab{}.
\newblock \showarticletitle{CAP twelve years later: How the "rules" have
  changed}.
\newblock \bibinfo{journal}{\emph{Computer}} \bibinfo{volume}{45},
  \bibinfo{number}{2} (\bibinfo{year}{2012}), \bibinfo{pages}{23--29}.
\newblock


\bibitem[\protect\citeauthoryear{Burckhardt}{Burckhardt}{2014}]%
        {Burckhardt14-book}
\bibfield{author}{\bibinfo{person}{Sebastian Burckhardt}.}
  \bibinfo{year}{2014}\natexlab{}.
\newblock \showarticletitle{Principles of Eventual Consistency}.
\newblock \bibinfo{journal}{\emph{Found. Trends Program. Lang.}}
  \bibinfo{volume}{1}, \bibinfo{number}{1–2} (\bibinfo{date}{Oct.}
  \bibinfo{year}{2014}), \bibinfo{pages}{1–150}.
\newblock
\showISSN{2325-1107}
\urldef\tempurl%
\url{https://doi.org/10.1561/2500000011}
\showDOI{\tempurl}


\bibitem[\protect\citeauthoryear{Burckhardt, Gotsman, Yang, and
  Zawirski}{Burckhardt et~al\mbox{.}}{2014}]%
        {Burckhardt14}
\bibfield{author}{\bibinfo{person}{Sebastian Burckhardt},
  \bibinfo{person}{Alexey Gotsman}, \bibinfo{person}{Hongseok Yang}, {and}
  \bibinfo{person}{Marek Zawirski}.} \bibinfo{year}{2014}\natexlab{}.
\newblock \showarticletitle{Replicated Data Types: Specification, Verification,
  Optimality}. In \bibinfo{booktitle}{\emph{Proceedings of the 41st ACM
  SIGPLAN-SIGACT Symposium on Principles of Programming Languages}} (San Diego,
  California, USA) \emph{(\bibinfo{series}{POPL '14})}.
  \bibinfo{publisher}{ACM}, \bibinfo{address}{New York, NY, USA},
  \bibinfo{pages}{271--284}.
\newblock
\showISBNx{978-1-4503-2544-8}
\urldef\tempurl%
\url{https://doi.org/10.1145/2535838.2535848}
\showDOI{\tempurl}


\bibitem[\protect\citeauthoryear{Emmi and Enea}{Emmi and Enea}{2018}]%
        {Emmi18}
\bibfield{author}{\bibinfo{person}{Michael Emmi} {and}
  \bibinfo{person}{Constantin Enea}.} \bibinfo{year}{2018}\natexlab{}.
\newblock \showarticletitle{Monitoring Weak Consistency}. In
  \bibinfo{booktitle}{\emph{Computer Aided Verification}},
  \bibfield{editor}{\bibinfo{person}{Hana Chockler} {and}
  \bibinfo{person}{Georg Weissenbacher}} (Eds.). \bibinfo{publisher}{Springer
  International Publishing}, \bibinfo{address}{Cham},
  \bibinfo{pages}{487--506}.
\newblock
\showISBNx{978-3-319-96145-3}


\bibitem[\protect\citeauthoryear{Emmi and Enea}{Emmi and Enea}{2019}]%
        {Emmi19}
\bibfield{author}{\bibinfo{person}{Michael Emmi} {and}
  \bibinfo{person}{Constantin Enea}.} \bibinfo{year}{2019}\natexlab{}.
\newblock \showarticletitle{Weak-consistency specification via visibility
  relaxation}.
\newblock \bibinfo{journal}{\emph{Proceedings of the ACM on Programming
  Languages}} \bibinfo{volume}{3}, \bibinfo{number}{POPL}
  (\bibinfo{year}{2019}), \bibinfo{pages}{1--28}.
\newblock


\bibitem[\protect\citeauthoryear{Enes, Almeida, Baquero, and Leitão}{Enes
  et~al\mbox{.}}{2019}]%
        {Enes19}
\bibfield{author}{\bibinfo{person}{Vitor Enes}, \bibinfo{person}{Paulo~Sérgio
  Almeida}, \bibinfo{person}{Carlos Baquero}, {and} \bibinfo{person}{João
  Leitão}.} \bibinfo{year}{2019}\natexlab{}.
\newblock \showarticletitle{Efficient Synchronization of State-Based CRDTs}. In
  \bibinfo{booktitle}{\emph{2019 IEEE 35th International Conference on Data
  Engineering (ICDE)}}. \bibinfo{pages}{148--159}.
\newblock
\urldef\tempurl%
\url{https://doi.org/10.1109/ICDE.2019.00022}
\showDOI{\tempurl}


\bibitem[\protect\citeauthoryear{Gibbons and Korach}{Gibbons and
  Korach}{1997}]%
        {Gibbons97}
\bibfield{author}{\bibinfo{person}{Phillip~B. Gibbons} {and}
  \bibinfo{person}{Ephraim Korach}.} \bibinfo{year}{1997}\natexlab{}.
\newblock \showarticletitle{Testing Shared Memories}.
\newblock \bibinfo{journal}{\emph{SIAM J. Comput.}} \bibinfo{volume}{26},
  \bibinfo{number}{4} (\bibinfo{date}{Aug.} \bibinfo{year}{1997}),
  \bibinfo{pages}{1208--1244}.
\newblock
\showISSN{0097-5397}
\urldef\tempurl%
\url{https://doi.org/10.1137/S0097539794279614}
\showDOI{\tempurl}


\bibitem[\protect\citeauthoryear{Gilbert and Lynch}{Gilbert and Lynch}{2012}]%
        {Gilbert12}
\bibfield{author}{\bibinfo{person}{Seth Gilbert} {and}
  \bibinfo{person}{Nancy~A. Lynch}.} \bibinfo{year}{2012}\natexlab{}.
\newblock \showarticletitle{Perspectives on the CAP Theorem}.
\newblock \bibinfo{journal}{\emph{Computer}} \bibinfo{volume}{45},
  \bibinfo{number}{2} (\bibinfo{year}{2012}), \bibinfo{pages}{30--36}.
\newblock
\urldef\tempurl%
\url{https://doi.org/10.1109/MC.2011.389}
\showDOI{\tempurl}


\bibitem[\protect\citeauthoryear{Golab, Li, and Shah}{Golab
  et~al\mbox{.}}{2011}]%
        {Golab11}
\bibfield{author}{\bibinfo{person}{Wojciech Golab}, \bibinfo{person}{Xiaozhou
  Li}, {and} \bibinfo{person}{Mehul~A. Shah}.} \bibinfo{year}{2011}\natexlab{}.
\newblock \showarticletitle{Analyzing consistency properties for fun and
  profit}. In \bibinfo{booktitle}{\emph{Proceedings of the 30th annual ACM
  SIGACT-SIGOPS symposium on Principles of distributed computing}}
  \emph{(\bibinfo{series}{PODC '11})}. \bibinfo{publisher}{ACM},
  \bibinfo{pages}{197--206}.
\newblock
\urldef\tempurl%
\url{http://doi.acm.org/10.1145/1993806.1993834}
\showURL{%
\tempurl}


\bibitem[\protect\citeauthoryear{Golab, Rahman, Auyoung, Keeton, and
  Gupta}{Golab et~al\mbox{.}}{2014}]%
        {Golab14}
\bibfield{author}{\bibinfo{person}{Wojciech Golab},
  \bibinfo{person}{Muntasir~Raihan Rahman}, \bibinfo{person}{Alvin Auyoung},
  \bibinfo{person}{Kimberly Keeton}, {and} \bibinfo{person}{Indranil Gupta}.}
  \bibinfo{year}{2014}\natexlab{}.
\newblock \showarticletitle{Client-Centric Benchmarking of Eventual Consistency
  for Cloud Storage Systems}. In \bibinfo{booktitle}{\emph{Proc. ICDCS}}.
  \bibinfo{pages}{493--502}.
\newblock


\bibitem[\protect\citeauthoryear{Golab, SteveLi, Lopez-Ortiz, and
  Nishimura}{Golab et~al\mbox{.}}{2018}]%
        {Golab18}
\bibfield{author}{\bibinfo{person}{W. Golab}, \bibinfo{person}{X. SteveLi},
  \bibinfo{person}{A. Lopez-Ortiz}, {and} \bibinfo{person}{N. Nishimura}.}
  \bibinfo{year}{2018}\natexlab{}.
\newblock \showarticletitle{Computing $k$-Atomicity in Polynomial Time}.
\newblock \bibinfo{journal}{\emph{SIAM J. Comput.}} \bibinfo{volume}{47},
  \bibinfo{number}{2} (\bibinfo{year}{2018}), \bibinfo{pages}{420--455}.
\newblock
\urldef\tempurl%
\url{https://doi.org/10.1137/16M1056389}
\showDOI{\tempurl}
\showeprint{https://doi.org/10.1137/16M1056389}


\bibitem[\protect\citeauthoryear{Jiang, Wei, and Huang}{Jiang
  et~al\mbox{.}}{2020}]%
        {Jiang20}
\bibfield{author}{\bibinfo{person}{Xue Jiang}, \bibinfo{person}{Hengfeng Wei},
  {and} \bibinfo{person}{Yu Huang}.} \bibinfo{year}{2020}\natexlab{}.
\newblock \showarticletitle{A Generic Specification Framework for Weakly
  Consistent Replicated Data Types}. In \bibinfo{booktitle}{\emph{2020
  International Symposium on Reliable Distributed Systems (SRDS)}}. IEEE,
  \bibinfo{pages}{143--154}.
\newblock


\bibitem[\protect\citeauthoryear{Lloyd, Freedman, Kaminsky, and Andersen}{Lloyd
  et~al\mbox{.}}{2013}]%
        {Lloyd13}
\bibfield{author}{\bibinfo{person}{Wyatt Lloyd}, \bibinfo{person}{Michael~J.
  Freedman}, \bibinfo{person}{Michael Kaminsky}, {and}
  \bibinfo{person}{David~G. Andersen}.} \bibinfo{year}{2013}\natexlab{}.
\newblock \showarticletitle{Stronger Semantics for Low-latency Geo-replicated
  Storage}. In \bibinfo{booktitle}{\emph{Proceedings of the 10th USENIX
  Conference on Networked Systems Design and Implementation}} (Lombard, IL)
  \emph{(\bibinfo{series}{nsdi'13})}. \bibinfo{publisher}{USENIX Association},
  \bibinfo{address}{Berkeley, CA, USA}, \bibinfo{pages}{313--328}.
\newblock
\urldef\tempurl%
\url{http://dl.acm.org/citation.cfm?id=2482626.2482657}
\showURL{%
\tempurl}


\bibitem[\protect\citeauthoryear{Motwani and Raghavan}{Motwani and
  Raghavan}{1995}]%
        {Motwani95}
\bibfield{author}{\bibinfo{person}{Rajeev Motwani} {and}
  \bibinfo{person}{Prabhakar Raghavan}.} \bibinfo{year}{1995}\natexlab{}.
\newblock \bibinfo{booktitle}{\emph{Randomized algorithms}}.
\newblock \bibinfo{publisher}{Cambridge university press}.
\newblock


\bibitem[\protect\citeauthoryear{Ouyang, Huang, Wei, and Lu}{Ouyang
  et~al\mbox{.}}{2021}]%
        {Ouyang21}
\bibfield{author}{\bibinfo{person}{Lingzhi Ouyang}, \bibinfo{person}{Yu Huang},
  \bibinfo{person}{Hengfeng Wei}, {and} \bibinfo{person}{Jian Lu}.}
  \bibinfo{year}{2021}\natexlab{}.
\newblock \showarticletitle{Achieving Probabilistic Atomicity With Well-Bounded
  Staleness and Low Read Latency in Distributed Datastores}.
\newblock \bibinfo{journal}{\emph{IEEE Transactions on Parallel and Distributed
  Systems}} \bibinfo{volume}{32}, \bibinfo{number}{4} (\bibinfo{year}{2021}),
  \bibinfo{pages}{815--829}.
\newblock
\urldef\tempurl%
\url{https://doi.org/10.1109/TPDS.2020.3034328}
\showDOI{\tempurl}


\bibitem[\protect\citeauthoryear{Shapiro, Pregui\c{c}a, Baquero, and
  Zawirski}{Shapiro et~al\mbox{.}}{2011}]%
        {Shapiro11a}
\bibfield{author}{\bibinfo{person}{Marc Shapiro}, \bibinfo{person}{Nuno
  Pregui\c{c}a}, \bibinfo{person}{Carlos Baquero}, {and} \bibinfo{person}{Marek
  Zawirski}.} \bibinfo{year}{2011}\natexlab{}.
\newblock \showarticletitle{Conflict-free Replicated Data Types}. In
  \bibinfo{booktitle}{\emph{Proceedings of the 13th International Conference on
  Stabilization, Safety, and Security of Distributed Systems}} (Grenoble,
  France) \emph{(\bibinfo{series}{SSS'11})}.
  \bibinfo{publisher}{Springer-Verlag}, \bibinfo{address}{Berlin, Heidelberg},
  \bibinfo{pages}{386--400}.
\newblock
\showISBNx{978-3-642-24549-7}
\urldef\tempurl%
\url{http://dl.acm.org/citation.cfm?id=2050613.2050642}
\showURL{%
\tempurl}


\bibitem[\protect\citeauthoryear{Wang, Enea, Mutluergil, and Petri}{Wang
  et~al\mbox{.}}{2019}]%
        {Wang19}
\bibfield{author}{\bibinfo{person}{Chao Wang}, \bibinfo{person}{Constantin
  Enea}, \bibinfo{person}{Suha~Orhun Mutluergil}, {and}
  \bibinfo{person}{Gustavo Petri}.} \bibinfo{year}{2019}\natexlab{}.
\newblock \showarticletitle{Replication-Aware Linearizability}. In
  \bibinfo{booktitle}{\emph{Proceedings of the 40th ACM SIGPLAN Conference on
  Programming Language Design and Implementation}} (Phoenix, AZ, USA)
  \emph{(\bibinfo{series}{PLDI 2019})}. \bibinfo{publisher}{Association for
  Computing Machinery}, \bibinfo{address}{New York, NY, USA},
  \bibinfo{pages}{980–993}.
\newblock
\showISBNx{9781450367127}
\urldef\tempurl%
\url{https://doi.org/10.1145/3314221.3314617}
\showDOI{\tempurl}


\bibitem[\protect\citeauthoryear{Wei, De~Biasi, Huang, Cao, and Lu}{Wei
  et~al\mbox{.}}{2016}]%
        {Wei16}
\bibfield{author}{\bibinfo{person}{Hengfeng Wei}, \bibinfo{person}{Marzio
  De~Biasi}, \bibinfo{person}{Yu Huang}, \bibinfo{person}{Jiannong Cao}, {and}
  \bibinfo{person}{Jian Lu}.} \bibinfo{year}{2016}\natexlab{}.
\newblock \showarticletitle{Verifying Pipelined-RAM Consistency over Read/Write
  Traces of Data Replicas}.
\newblock \bibinfo{journal}{\emph{IEEE Trans. Parallel Distrib. Syst.}}
  \bibinfo{volume}{27}, \bibinfo{number}{5} (\bibinfo{date}{May}
  \bibinfo{year}{2016}), \bibinfo{pages}{1511--1523}.
\newblock
\showISSN{1045-9219}
\urldef\tempurl%
\url{https://doi.org/10.1109/TPDS.2015.2453985}
\showDOI{\tempurl}


\bibitem[\protect\citeauthoryear{Wei, Huang, and Lu}{Wei et~al\mbox{.}}{2017}]%
        {Wei17}
\bibfield{author}{\bibinfo{person}{Hengfeng Wei}, \bibinfo{person}{Yu Huang},
  {and} \bibinfo{person}{Jian Lu}.} \bibinfo{year}{2017}\natexlab{}.
\newblock \showarticletitle{Probabilistically-Atomic 2-Atomicity: Enabling
  Almost Strong Consistency in Distributed Storage Systems}.
\newblock \bibinfo{journal}{\emph{IEEE Trans. Comput.}} \bibinfo{volume}{66},
  \bibinfo{number}{3} (\bibinfo{date}{March} \bibinfo{year}{2017}),
  \bibinfo{pages}{502--514}.
\newblock
\showISSN{0018-9340}
\urldef\tempurl%
\url{https://doi.org/10.1109/TC.2016.2601322}
\showDOI{\tempurl}


\bibitem[\protect\citeauthoryear{Yu and Vahdat}{Yu and Vahdat}{2002}]%
        {Yu02-TOCS}
\bibfield{author}{\bibinfo{person}{Haifeng Yu} {and} \bibinfo{person}{Amin
  Vahdat}.} \bibinfo{year}{2002}\natexlab{}.
\newblock \showarticletitle{Design and Evaluation of a Conit-based Continuous
  Consistency Model for Replicated Services}.
\newblock \bibinfo{journal}{\emph{ACM Trans. Comput. Syst.}}
  \bibinfo{volume}{20}, \bibinfo{number}{3} (\bibinfo{date}{Aug.}
  \bibinfo{year}{2002}), \bibinfo{pages}{239--282}.
\newblock
\showISSN{0734-2071}
\urldef\tempurl%
\url{https://doi.org/10.1145/566340.566342}
\showDOI{\tempurl}


\bibitem[\protect\citeauthoryear{Zhang, Wei, and Huang}{Zhang
  et~al\mbox{.}}{2021}]%
        {Zhang21}
\bibfield{author}{\bibinfo{person}{Yuqi Zhang}, \bibinfo{person}{Hengfeng Wei},
  {and} \bibinfo{person}{Yu Huang}.} \bibinfo{year}{2021}\natexlab{}.
\newblock \showarticletitle{Remove-Win: a Design Framework for Conflict-free
  Replicated Data Types}. In \bibinfo{booktitle}{\emph{Proceedings of the IEEE
  International Conference on Parallel and Distributed Systems}}
  \emph{(\bibinfo{series}{ICPADS'21})}. \bibinfo{publisher}{IEEE},
  \bibinfo{address}{Beijing, China}.
\newblock


\end{thebibliography}

\end{CJK*}

\end{document}